\documentclass[preprint,12pt]{elsarticle}

\usepackage{graphicx}

\usepackage{amssymb}
\usepackage{amsthm, amsmath}

\usepackage{xcolor}

\usepackage{verbatim}

\journal{ArXiv}

\begin{document}

\begin{frontmatter}



\title{Salt-induced reentrant stability of polyion-decorated particles with tunable surface charge density}


\author[a,b]{Simona Sennato}
\author[b]{Laura Carlini}
\author[c]{Domenico Truzzolillo}
\author[b]{Federico Bordi}

\address[a]{CNR-ISC UOS Roma- c/o Dipartimento di Fisica - Sapienza Universit\`{a} di Roma - P.zzle A. Moro, 2 - 00185 Roma - Italy}
\address[b]{Dipartimento di Fisica - Sapienza Universit\`{a} di Roma - P.zzle A. Moro, 2 - 00185 Roma - Italy}
\address[c]{Laboratoire Charles Coulomb (L2C), UMR 5221 CNRS-Universit\'{e} de Montpellier,
Montpellier, France}

\begin{abstract}

The electrostatic complexation between DOTAP-DOPC unilamellar liposomes and an oppositely charged polyelectrolyte (NaPA) has been investigated in a wide range of the liposome surface charge density. We systematically characterized the "reentrant condensation" and the charge inversion of polyelectrolyte-decorated liposomes by means of dynamic light scattering and electrophoresis. We explored the stability of this model polyelectrolyte/colloid system by fixing each time the charge of the bare liposomes and by changing two independent control parameters of the suspensions: the polyelectrolyte/colloid charge ratio and the ionic strength of the aqueous suspending medium. The progressive addition of neutral DOPC lipid within the liposome membrane gave rise to a new intriguing phenomenon: the stability diagram of the suspensions showed a \emph{novel reentrance} due to the crossing of the desorption threshold of the polyelectrolyte. Indeed, at fixed charge density of the bare DOTAP/DOPC liposomes and for a wide range of polyion concentrations, we showed that the simple electrolyte addition first (low salt regime) destabilizes the suspensions because of the enhanced screening of the residual repulsion between the complexes, and then (high salt regime) determines the onset of a new stable phase, originated by the absence of polyelectrolyte adsorption on the particle surfaces.
We show that the observed phenomenology can be rationalized within the Velegol-Thwar model for heterogeneously charged particles and that the polyelectrolyte desorption fits well the predictions of the adsorption theory of Winkler and Cherstvy \cite{Winkler06}.
Our findings unambiguously support the picture of the reentrant condensation as driven by the correlated adsorption of the polyelectrolyte chains on the particle surface, providing interesting insights into possible mechanisms for tailoring complex colloids via salt-induced effects.\\
\end{abstract}


\end{frontmatter}

\section{Introduction}
The phenomenon of complexation between oppositely charged colloids or macromolecules in aqueous solution is a fundamental process, widely exploited by nature, as for example in DNA packaging within biological cells \cite{Volkenstein}, but also by technology, as in industrial colloidal stabilization, water treatment and paper making \cite{Napper}.\\
Being the result of a delicate balance of forces of different nature, small variations in the physico-chemical parameters may induce large changes in the resulting complexes  \cite{amenitsch2011existence}. Electrostatics is clearly the relevant interaction in driving the aggregation, although non-electrostatic interactions can be important in modulating the process and can deeply affect the characteristics of the resulting self-assembled structures. However, due to the great complexity of these systems, also
 the details of purely electrostatic interactions remain to be completely clarified, despite the intense experimental, theoretical and computational research aimed to understand the mechanisms driving the formation and the stabilization of the complexes \cite{levin2002electrostatic,Borkovec14p03} in the physical context of macroion-multivalent counterion interactions \cite{nguyen2000macroions}.\\
The possibility to get a fine control of the stability of the complexes by tuning the competition between attractive and repulsive electrostatic interactions justifies the actual enduring attention on polyion-colloid complexation, being key to the technological development of nano-structured materials or nano-devices to be used for example in drug delivery or gene-therapy \cite{Agrati10,Bordi14p184}.\\
In particular, a rich literature exists on the variety of structures formed by charged liposomes interacting via electrostatic forces with oppositely charged linear polyions, whether they are synthetic polymers, polypeptides or DNA \cite{Bordi09Review,Quemeneur10,safinya2001structures}.
A feature characterizing the phase behavior of these systems is the presence of a "reentrant condensation" accompanied by a marked "overcharging", or charge inversion \cite{Bordi04,Sennato05p11}. Indeed, when a given volume of a liposome suspension is mixed with the same volume of a solution containing oppositely charged polyelectrolytes, the rapid formation of clusters is systematically observed. For low enough polyelectrolyte concentration the clusters are small (they are formed by a few liposomes), and after their rapid formation they are very stable. The measured $\zeta$-potential of these clusters is only slightly lower, in absolute value, than that measured for the liposomes in 
 absence of adsorbed polyelectrolytes. As the polyion concentration (and hence the ratio, $\xi$, between the number of stoichiometric charges on the polymer and on the particles) is increased, larger but still stable clusters are formed, with a $\zeta$-potential which is further reduced in absolute value. At a sufficiently high polyelectrolyte concentration, the liposome suspension is completely destabilized, the large clusters that form are not stable, their $\zeta$-potential is close to zero and they rapidly coalesce in macroscopic "flocs". Going beyond the \emph{isoelectric point}, hence further increasing the polyelectrolyte concentration, stable clusters form again, but now their size \emph{decreases} upon increasing the polyelectrolyte concentration in the mixture: the polyelectrolyte-induced condensation of the liposomes is hence "reentrant". At the same time, the $\zeta$-potential after having passed through the zero, increases again in absolute value, and the sign of the excess charge of the clusters is that of the polyelectrolyte, i.e. opposite to that of the original liposomes (charge inversion).\\
As the polyelectrolyte concentration is even further increased some "free" polyelectrolyte (non-complexed) chains begin to coexist with the overcharged small clusters or single liposomes. Eventually, at even higher polyelectrolyte content and due to the strong electrostatic screening, the colloidal suspension is completely destabilized, as expected from the Derjaguin-Landau-Verwey-Overbeek (DLVO) theory \cite{Sennato04p296}.\\
It can be shown \cite{Grosberg02} that both the overcharging and the reentrant condensation are consequences of the \emph{correlated} adsorption of polyelectrolyte chains on the particle surface. Indeed, the electrostatic attraction of the polyelectrolytes to the oppositely charged particle surface, the repulsion between the like-charged adsorbed chains and their confinement entropy \cite{Dobrynin00,Grosberg02,Bordi09Review, Truzzolillo10} compete to give rise to adsorption patterns that are characterized by long range correlations. This correlated adsorption is also a consequence of the large difference between the diffusivities of the colloidal particles and the polyelectrolyte chains \cite{schonhoff2003layered} that reflects in different timescales of adsorption and aggregation. When the polyelectrolyte and the oppositely charged colloidal particles are mixed together, the adsorption of the polyelectrolyte on the particle surface is a much faster process than the aggregation. The huge surface/volume ratio which characterizes the colloid, and the distribution of this surface within the whole volume of the host phase, speed up
dramatically the adsorption process, decreasing the time  needed by
for the polyelectrolyte chains to reach the adsorbing surface by diffusion. Conversely, the aggregation, being controlled by the diffusivity of the bulkier colloidal particles, occurs on longer timescales \cite{Volodkin07}. As a result, when the concentration of the polyelectrolyte is not enough to "saturate" the available surface of the particles, the latter appears "decorated" by statistically regular patterns rather than uniformly coated by the polyelectrolytes \cite{Sennato05p11}. This alternation of domains of "polyelectrolyte-covered" and "bare" particle surface results in a non-uniform surface charge distribution \cite{Borkovec14p03}. Such a "patchy" distribution of charge, together with the screening effect given by the small counterions, generates a short range attraction between these "polyelectrolyte-decorated particles" (hereafter pd-particles), that has been widely studied in the last two decades \cite{Miklavic94,Khachatourian98,Velegol01,Mukherjee04}. The superposition of this attraction, the ubiquitous van der Waals interaction and the electrostatic repulsion due to the residual net charge of the complexes, results in a inter-particle potential characterized by a barrier that, in proper conditions, yields the formation of stable, finite size clusters \cite{Sennato08}, even if all the primary particles bear the same net charge \cite{Truzzolillo10}, the adsorbed polyelectrolyte layer representing a sort of electrostatic glue that sticks together the particles \cite{Bordi05p134}. Noteworthy, this mechanism works in principle not only for liposomes, but for whatever colloid and polyelectrolyte, as long as the proper "patchy" charge distribution can be established. In fact in the presence of an oppositely charged polyelectrolyte many colloids, characterized by different structures and chemical compositions, like bare \cite{Keren02} and lipid-coated latex particles \cite{Zuzzi08}, ferric oxide particles \cite{Radeva08}, dendrimers \cite{Kabanov00} and micelles \cite{Wang00}, show the "reentrant condensation" phenomenon.\\

In a recent paper \cite{SennatoSM2012}, we have systematically explored the phase diagram and the reentrant condensation of cationic DOTAP liposomes in the presence of the anionic polyelectrolyte sodium poly\-acrylate (NaPA) and at varying the ionic strength of the aqueous suspending medium. In this work we focus instead on the effect on the phase diagram of the liposome surface charge density. To this aim, to vary the surface charge, we employed liposomes prepared from mixtures of a cationic lipid, Di-oleoyl-trimethyl-ammonium-propane [DOTAP],  and a zwitterionic phospholipid, Di-oleoyl-phosphatidyl-coline [DOPC], at different molar fraction of the two components. The reentrant condensation of these mixed DOTAP-DOPC pd-liposomes was explored by varying the ionic strength of the solution with the addition of proper amounts of NaCl.\\
DOTAP-DOPC-NaPA is a good model system to study the complex but general phenomenology that appears by mixing oppositely charged colloidal particles and polyelectrolytes. Moreover, due to the good biocompatibility of NaPA \cite{Falk01,DeGiglio10}, the DOTAP-DOPC-NaPA complexes have also a specific interest for their potential use as multi-compartment vectors for multi-drug delivery \cite{Agrati10}. A deeper understanding of the relation between polyion adsorption, pd-liposome interactions and stability of the resulting aggregates is key for further developments of functional colloidal particles with application in biotechnology and material science.

\section{Experimental}
\subsection{Materials}
The cationic lipid 1,2-dioleoyl-3-trimethylammonium-propane (chloride salt) [DOTAP] and the zwitterionic phospholipid 1,2-dioleoyl-sn-glycero-3-phosphocholine [DOPC] were  purchased from Avanti Polar Lipids (Alabaster, AL) and used without further purification. The negatively charged polyelectrolyte sodium poly-acrylate,  $[-CH_2CH(CO_2Na)-]_n$ [NaPA], with nominal molecular weight 60 kD, was purchased from Polysciences Inc. (Warrington, PA) as 25 $\%$ aqueous solution. All liposome samples and polyelectrolyte solutions were prepared in Milli-Q  grade water, with electrical conductivity less than 1$\cdot$10$^{-6}$ S/cm. Sodium Chloride (analytical grade) was from Merck (Germany).

\subsection{Preparation of cationic lipid-polyion complexes}
DOTAP/DOPC mixed liposomes were prepared by dissolving appropriate amounts of the two lipids in methanol-chloroform solution (1:1 vol/vol). In all the experiments the total lipid concentration was kept constant at approximately 1.6 mg/ml, while  the molar fraction of the cationic DOTAP, $X = n_{DOTAP}/(n_{DOTAP}+n_{DOPC})$  (where $n_{i}$ is the number of moles of the specie $i$)  was varied  from 0.2 to 0.75. The solvent was evaporated by overnight roto-evaporation in vacuum. The dried lipid film was then re-hydrated with Milli-Q quality water, for 1 hour at a temperature of $40^{\circ}$C, well above the main phase transition temperature, $T_m$, of both DOPC and DOTAP ($\approx -17 ^{\circ}$C and $ \approx 0 ^{\circ}$C, respectively). In order to form small uni-lamellar vesicles, the lipid aqueous suspension was sonicated for 1 hour at a pulsed power mode until the solution appeared optically transparent in white light. Finally, the vesicle dispersion was filtered using a Millipore 0.4 $\mu m$ polycarbonate filter. The liposome dispersions were stored at $4^{\circ}$C.\\
For all the liposome preparations, we checked the size distribution by Dynamic Light Scattering (DLS). The distribution was log-normal with a mean hydrodynamic radius of $40\pm 5$ nm and a polydispersity of the order of $0.2 \div 0.3$.\\

Each liposome-polyelectrolyte mixture was prepared according to the following standard procedure \cite{Bordi04}: a volume of 0.4 ml of the polyelectrolyte solution, at the required concentration and ionic strength, was added to an equal volume of the liposome dispersion in a single mixing step and gently agitated by hand. Before mixing, the liposome suspension and polyelectrolyte solution were kept in a thermostatted bath at the temperature of the experiment, to avoid interference of thermal gradients during the following measurement. After mixing the two components, the sample was immediately placed in the thermostatted cell holder of the instrument for the measurement of the electrophoretic mobility, the size and the size distribution of the resulting aggregates. Electrophoretic mobility measurements were systematically performed 5 minutes after the mixing, immediately followed by the size determination. During the experiments the temperature was controlled within $\pm 0.1^{\circ}$C. \\
The ionic strength due to the 1-valent counterions Na$^+$ and Cl$^-$ in the pd-liposome suspensions was varied from 0.005 M to 0.8 M by adding appropriate amounts of NaCl to the polyelectrolyte solution to be mixed with the liposome suspensions.

\subsection{Dynamic light scattering measurements}

Size and size distribution of liposomes and polyions-liposome aggregates were characterized by means of dynamic light scattering (DLS) measurements. For light-scattering measurements, a MALVERN NanoZetasizer apparatus equipped with a 5 mW HeNe laser was employed (Malvern Instruments LTD, UK). This system uses  backscatter detection, i.e. the scattered light is collected at an angle of 173$^o$.  The main advantage of this detection geometry, when compared to the more conventional 90$^o$, is its inherent larger insensitiveness to multiple scattering effects  \cite{Dhadwal91}. Intuitively, since nor the illuminating laser beam, nor the detected scattered light need to travel through the entire sample, the chance that incident and scattered photons will encounter more than one particle is reduced.  Moreover, as large particles scatter mainly in the forward direction, the effects of dust or, as is our case, of large irregular aggregates (lumps or clots) on the size distribution are greatly reduced.\\
To obtain the size distribution, the measured autocorrelation functions were analyzed by means of the CONTIN algorithm \cite{Provencher82}. Decay times are used to determine the distribution of the diffusion coefficients $D$ of the particles, which in turn can be converted in a distribution of apparent hydrodynamic radii, $R_H$, using the Stokes$-$Einstein relationship $R_H = K_B T/6 \pi \eta D$, where $K_BT$ is the thermal energy and $\eta$ the solvent viscosity. The values of the radii shown here correspond to the average values on several measurement and are obtained from intensity weighted distributions \cite{Provencher82,Devos96}.

\subsection{Electrophoretic mobility measurements}

The electrophoretic mobility of the suspended particles was measured by means of the same NanoZetaSizer apparatus employed for DLS measurements. This instrument is integrated with a laser Doppler electrophoresis technique, and the particle size and electrophoretic mobility can be measured almost simultaneously and in the same cuvette. In this way, possible experimental uncertainties due to different sample preparations, thermal gradients and convection are significantly reduced.\\
Electrophoretic mobility is determined using the Phase Analysis Light Scattering (PALS) technique \cite{Tscharnuter01}, a method which is especially useful at high ionic strengths, where mobilities are usually low. In these cases the PALS configuration has been shown to be able to measure mobilities two orders of magnitudes lower than traditional light scattering methods based on the shifted frequency spectrum (spectral analysis).\\
In general, the conversion of the electrophoretic mobility into the $\zeta$-potential for a colloidal particle is not a trivial matter, because it involves a rather complex hydrodynamical problem \cite{Hidalgo96}. However, if the reduced electrokinetic radius $\kappa R$ (with $\kappa$ the inverse Debye's screening length and $R$ the particle's radius), is large enough, the $\zeta$-potential can be straightforwardly obtained from the Helmholtz/Smoluchowski (HS) approximation:
\begin{equation}\label{eq:zetapot}
 \mu=\varepsilon_0 \varepsilon_r \zeta / \eta
\end{equation}
where $\epsilon_0$ and $\epsilon$ are the vacuum and solvent permittivity, respectively.

\section{Theoretical and phenomenological background}

\begin{figure}[!ht]
  \includegraphics[width=8 cm]{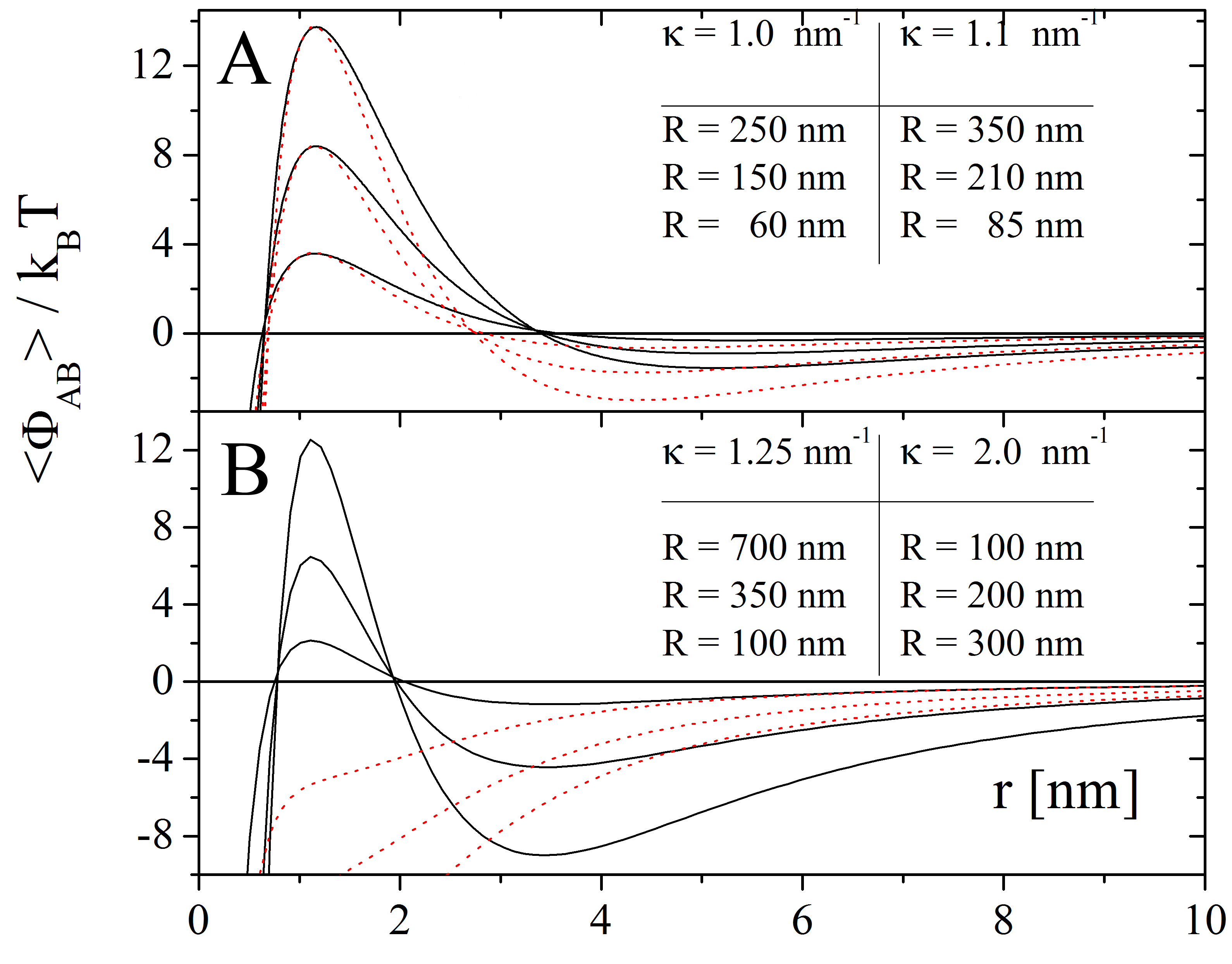}\\
  \caption{(Color online) The potential of mean force $\langle\Phi_{AB}(r)\rangle$ between the pd-liposomes (equation (\ref{eq:potential})) in units of the thermal energy $k_B T$ as a function of the distance $r$ between the particle's surfaces, calculated for typical values of the surface potential average value, $\Psi_A=\Psi_B = 25\, mV$, and standard deviation, $\sigma_A=\sigma_B= 15\, mV$, at $T= 298.16 \, K^\circ$, with the Hamaker constant $\mathcal{H} = 10^{-20} \, J $  and with a thickness of the liposome bilayer  $h_A=h_B= 5 \, nm$. In panel A the curves are calculated at two different values of the ionic strength and hence of the electrostatic screening, i.e. at $\kappa= 1\,nm^{-1}$ (full lines) and  $\kappa= 1.1 \,nm^{-1}$ (dotted lines) and for several different values of the radius $R$ of the particles, to show that at increasing the screening, a same height of the potential barrier is reached for a larger radius of the aggregates. As the salt content is increased, panel B ($\kappa= 1.25 \,nm^{-1}$, full lines), the secondary minimum in front of the potential barrier may become deep enough to destabilize the suspension. As expected, at much higher ionic strengths ($\kappa= 2.0 \,nm^{-1}$, dotted lines) the repulsive barrier disappears, there is only one deep minimum and the colloid dispersion is completely destabilized.} \label{Fig:Potential}
\end{figure}

Monte Carlo simulations have recently shown that the reentrant condensation of pd-particles can be rather satisfactorily described by assuming that two decorated particles ($A$ and $B$) interact via the potential of mean force \cite{Truzzolillo09,Truzzolillo10,SennatoSM2012}
\begin{equation}\label{eq:potential}
\begin{split}
\langle\Phi_{AB}(r)\rangle = \frac{ R_AR_B}{R_A+R_B} \Biggl \{ \varepsilon_0 \varepsilon_r
\Biggl [ (\Psi_A^2+\Psi_B^2+\sigma_A^2+\sigma_B^2)\ln(1-e^{-2\kappa r} )+\\
  2\Psi_A\Psi_B\ln\left(\coth\frac{\kappa r}{2}\right) \Biggr]
  -\frac{\mathcal{H}}{6} \Biggl ( \frac{1}{r+h_A+h_B}-\frac{1}{r+h_A}-\frac{1}{r+h_B}+\frac{1}{r} \Biggr ) \Biggr \} \\ -\frac{\mathcal{H}}{6}\ln \left[\frac{r(r+h_A+h_B)}{(r+h_A)(r+h_B)}\right].
\end{split}
\end{equation}
Here $R_i$ is the radius of the particle $i$ (with $i\equiv A,B$), $r$ is the minimal distance between the two interacting particle surfaces, and $\Psi_i$ and $\sigma_i$ are the values of the mean and standard deviation of the surface potential of the particles (assumed spherical).\\
The last two terms, proportional to the Hamaker constant $\mathcal{H}$, take into account the van der Waals forces (between shelled spheres with radii $R_i$ and shell thickness $h_i$ \cite{Tadmor01}), while the terms within the square brackets represent the screened electrostatic interaction between two inhomogeneously charged spheres in an electrolyte solution, taking into account the presence of the ionic double layer at their surface \cite{Velegol01}. Such potential corresponds to an extension of the classical Hogg-Healy-Fuerstenau (HHF) theory for homogeneously charged spheres \cite{Hogg66} to which reduces for $\sigma_A=\sigma_B = 0$.\\
Both the electrostatic and the van der Waals contributions in eq. (\ref{eq:potential}) are calculated assuming the validity of the Derjaguin approximation \cite{Derjaguin34,Israelachvili85}. Within this approximation, the force between the surfaces of two interacting particles is proportional to an effective radius, whose reciprocal is the arithmetic mean of the inverse curvature radii of the surfaces involved, $R_A$ and $R_B$ (the prefactor in front of the curly brackets in eq. (\ref{eq:potential})). The Derjaguin approximation holds for inter-particle distances $d \ll R_A$, $R_B$ and for $\kappa^{-1} \ll R_A$, $R_A$. In these conditions, the approximation has been recently investigated experimentally and found "surprisingly robust" \cite{Borkovec06}.\\
It is easy to see that the first electrostatic term  in eq. (\ref{eq:potential}) is always attractive (negative) also for like-charged particles. The presence of this attractive term is consistent with the results of Poisson-Boltzmann equation (PBE) approaches and simulations \cite{Dzubiella03,Granfeldt91,Truzzolillo10}, pointing out the existence of a short range attraction between polyelectrolyte-decorated macroions. The term "short range" is well justified since the decaying length of the attractive term $(2\kappa)^{-1}$ is one half of the Debye screening length, that sets the scale for the screened electrostatic interactions.\\
Figure (\ref{Fig:Potential}) shows the typical behavior of the potential of mean force $\langle\Phi_{AB}(r)\rangle$ as a function of the distance $r$ between the particle surfaces. The values of the potential, in units of the thermal energy $k_B T$, are calculated for typical values of the average surface potential, $\Psi_A=\Psi_B = 25\, mV$, and standard deviation, $\sigma_A=\sigma_B= 15\, mV$, at temperature $T= 298.16 \, K^\circ$, with the Hamaker constant $\mathcal{H} = 10^{-20} \, J $  and for a thickness of the liposome bilayer  $h_A=h_B= 5 \, nm$. The curves are calculated at three different ionic strengths and for several different values of the particle's radius $R$. At a given value of the ionic strength and hence of the electrostatic screening length, $\kappa^{-1}$, the height of the potential barrier increases with the Derjaguin effective radius $R$ of the aggregates \cite{Truzzolillo09}.\\
Within this framework, the aggregation of the pd-particles can hence be described as a thermally activated process \cite{Sennato08} where, since the potential barrier height increases with their own size, a finite size of the clusters turns out to be eventually stable \cite{Truzzolillo09,Sennato09}.\\
Panel A of figure (\ref{Fig:Potential}) shows how the ionic strength tunes the size of the stable clusters: as the salt concentration is increased a larger size of the aggregates is needed in order to get the same barrier profile.\\
At much higher ionic strengths (Panel B of figure (\ref{Fig:Potential}), dotted lines) the repulsive barrier disappears, there is only one deep global minimum, and the colloid dispersion is completely destabilized. However, already at intermediate values of the salt content (full lines in panel B) the secondary minimum in front of the barrier begins to deepen enough to destabilize the dispersion.\\

\begin{figure}[!ht]
  \includegraphics[width=8 cm]{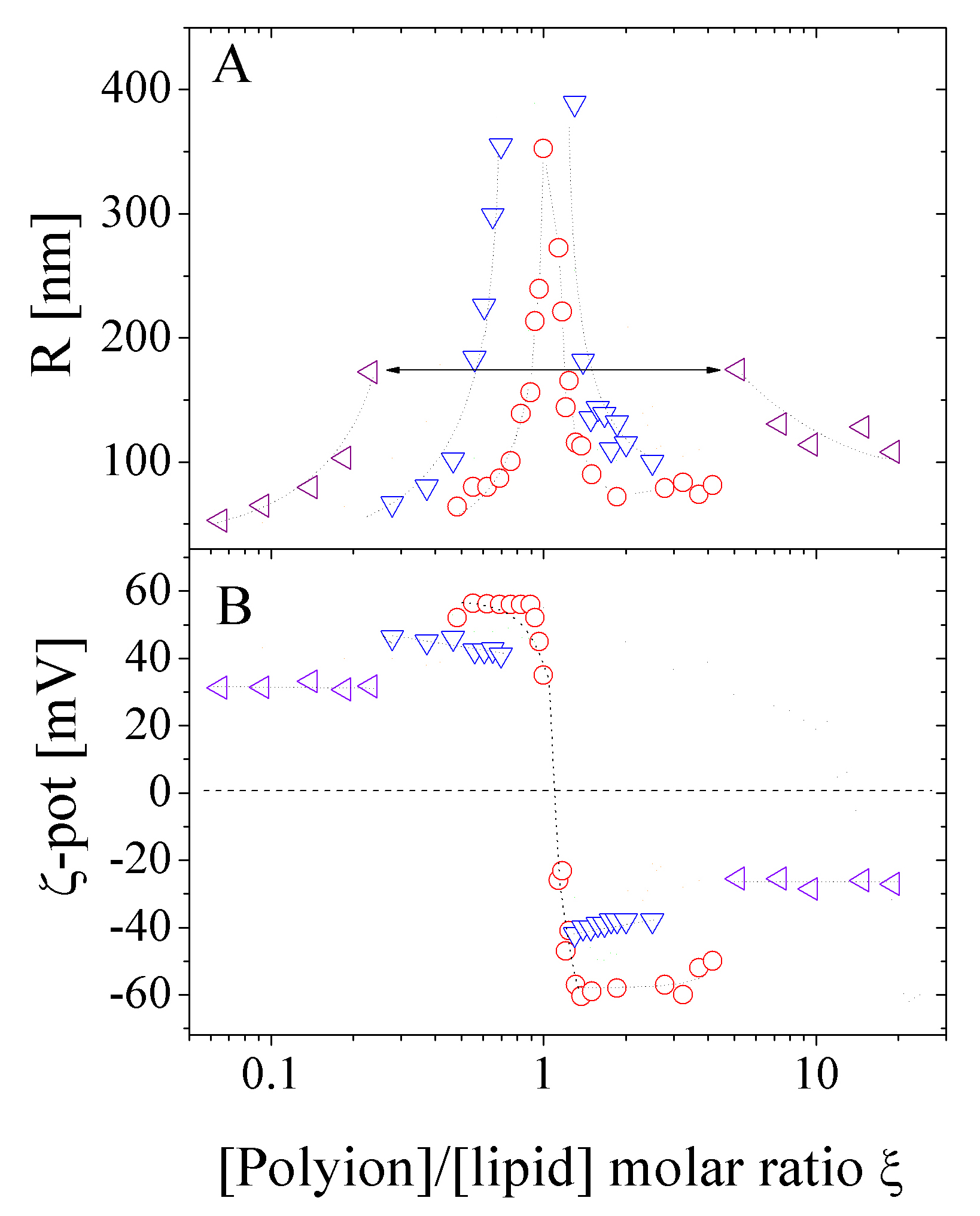}\\
  \caption{(Color online) The average hydrodynamic radius (panel A) and the corresponding $\zeta$-potential (panel B) of polyion-liposome clusters as a function of the polyion/lipid charge ratio $\xi$, in the presence of different amounts of a simple electrolyte, NaCl, added to a final concentration of: 0.4 M  ($\lhd$);  0.2 M   ($\triangledown$);   0.005 M ($\circ$). Temperature was fixed at 25 $^{\circ}$C. As an example, the arrow spans the instability range of the 0.4 M NaCl sample. Adapted with changes from \cite{SennatoSM2012}.}\label{Fig:PZDOTAP}
\end{figure}

This theoretical picture is in excellent agreement with the observed phenomenology, as we will further point out later. Figure (\ref{Fig:PZDOTAP}) shows the typical reentrant condensation behavior of pure DOTAP/NaPA decorated liposomes in the presence of NaCl. Consistently with the prediction of the theory, for a given value of the polyelectrolyte/lipid charge ratio $\xi$ the size of the aggregates increases with the salt concentration. The symbols in the figure corresponds to stable aggregates, i.e. aggregates whose size, once formed (on a much shorter timescale than the measurement time, which is typically a few minutes), remains constant for several weeks. At the lowest concentration of the electrolyte the suspension is unstable only in a very narrow zone close to the isoelectric condition and the reentrant condensation curve appears almost continuous. Indeed, it may be even difficult to locate experimentally the precise polyelectrolyte/lipid charge ratio that results in instability and phase separation. However, as the electrolyte concentration is further increased, an ample gap of instability appears around the isoelectric point, where stable aggregates does not form and whose width increases with the salt content.  Within this range of polyelectrolyte/lipid charge ratios, as the polyelectrolyte solution and the liposome suspension are mixed together, large  mesoscopic "flocs" immediately begins to form, that rapidly coalesce and eventually separate from the aqueous solution, floating on its surface (flocculation).\\
These features appear to be rather general. A similar behavior was observed, for example, also in the aggregation of oppositely charged polyelectrolytes and micelles \cite{Wang99}. In that system, consistently with the reentrant condensation scenario, an increasing salt concentration causes the onset and further expansion of coacervation phenomena close to the isoelectric point.\\

\begin{figure}[!ht]
  \includegraphics[width=8cm]{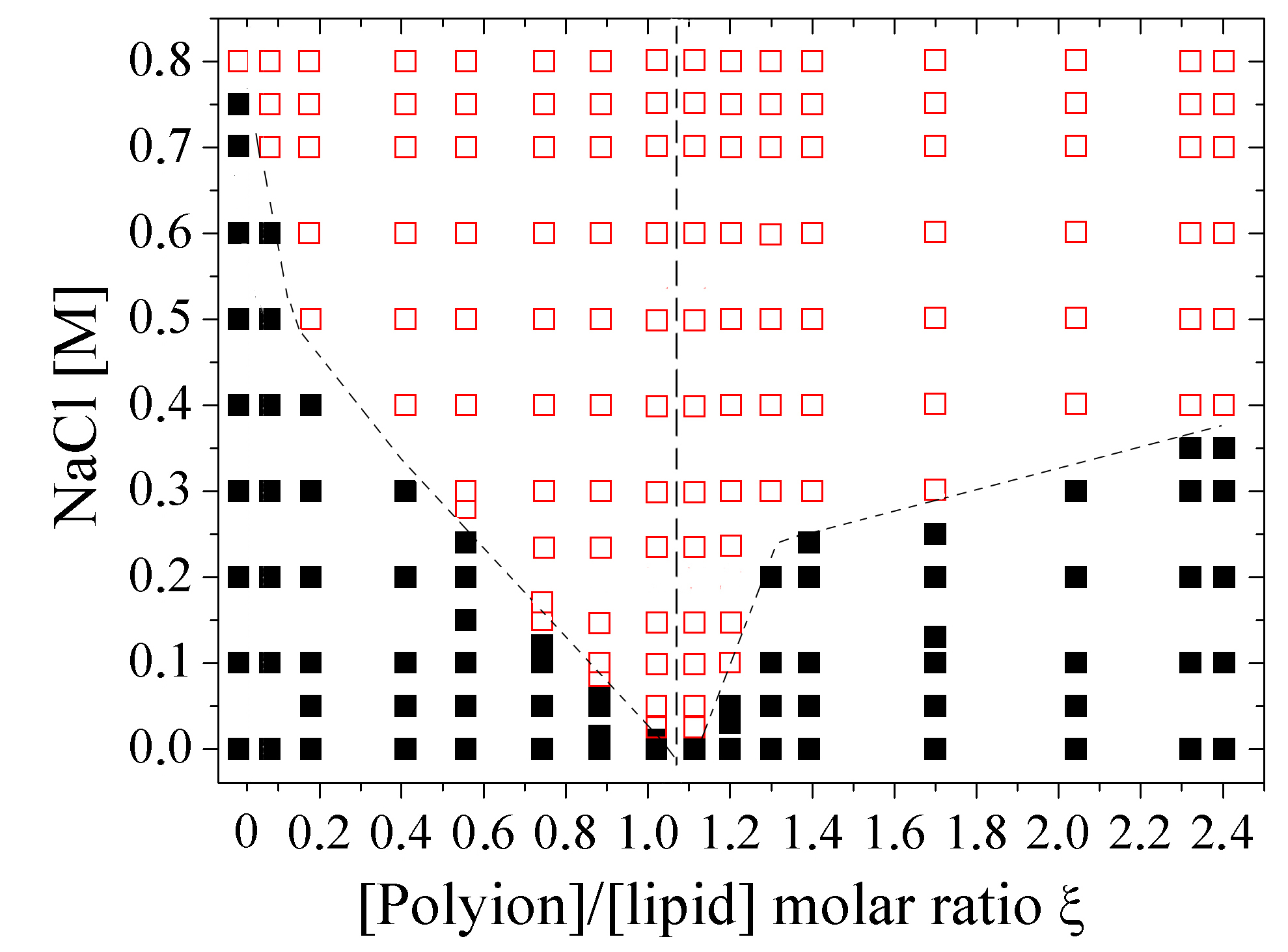}\\
  \caption{(Color online) Phase diagram of the pd-liposome complexes formed by pure DOTAP liposomes in the presence of added salt. At each given polyion/lipid molar ratio $\xi$, and for a defined concentration of NaCl added, full symbols represent stable samples, open symbols represent unstable samples (data partially published in \cite{SennatoSM2012}).}\label{fig:DFDOTAP}
\end{figure}

The behavior of the $\zeta$-potential of the pd-liposome aggregates is shown in Panel B of fig. \ref{Fig:PZDOTAP}. As expected from the Gouy-Chapman-Stern theory \cite{Cevc93,Israelachvili85}, the absolute value of the $\zeta$-potential decreases as the salt content increases both below and above the isoelectric point and for a given polyelectrolyte-lipid charge ratio. Beyond the isoelectric point the $\zeta$-potential decreases even more markedly (in absolute value) with the increasing salt content. This effect can be interpreted within the framework of the "screening-reduced" adsorption regime described by Muthukumar \cite{Muthukumar87}. This behavior has to be expected when in the adsorption of a highly charged polyelectrolyte on an oppositely charged surface the driving attraction is mainly electrostatic, and the added salt screens equally well the chain–chain repulsion and the chain–surface attraction. Evidence of this regime can be also found in the MC simulations of Truzzolillo et al. \cite{Truzzolillo10}, where the dependence of the thickness of the polyelectrolyte layer adsorbed on single pd-particles is investigated as a function of the ionic strength of the solution.\\
Fig. \ref{fig:DFDOTAP} shows the NaCl-$\xi$ phase diagram of pd-liposomes complexes formed by pure DOTAP liposomes. Stable mesoscopic aggregates or unstable large clusters, which eventually flocculate, are formed depending on $\xi$ and on the amount of the added salt (plotted in full and open symbols, respectively). The position of the isoelectric point is marked by a vertical dashed line. Below the isoelectric point, the boundary line between the stable and the unstable regions is approximately straight, on the contrary once the isoelectric point is crossed, there is a clear change of the slope of the boundary line at $\xi\approx 1.3$. This inflection point corresponds to the maximum overcharging of the liposomes: after this point the particles surface is completely saturated by the polyelectrolyte and the "stable aggregates" correspond to single liposomes coated by an almost homogeneous layer of polyelectrolyte. Beyond the inflection point ($\xi\approx 1.3$) a further increase of the polyelectrolyte concentration turn out simply into an increase of the "free polyelectrolyte" amount in solution, and does not affect the net charge of the complexes. The cause of this decrease of the boundary slope  is that in this regime, also the free polyelectrolyte (and its counterions) contribute to the electrostatic screening, so that a comparatively lower NaCl concentration produces a similar destabilizing effect on the suspension \cite{Sennato04p296}. The complexation experiments have been performed increasing the NaCl concentration up to 1 M.  Above 0.6 M NaCl the colloidal suspension is completely destabilized also at the lowest polyelectrolyte/lipid ratio investigated, and above 0.8 M NaCl even in the absence of the polyelectrolyte (see also  \cite{SennatoEPL04}).

\section{Results}

We focused here on the effect of the surface charge density of the liposomes on the stability/instability diagram of pd-liposomes as a function of the polyelectrolyte/lipid charge ratio $\xi$ and of the ionic strength of the solution.
To this aim we prepared DOTAP/DOPC liposomes at different molar fraction $X$ of the charged lipid and measured the hydrodynamic radius and the electrophoretic mobility of the resulting pd-liposome complexes, varying the polyelectrolyte/lipid charge ratio $\xi$ by changing the polyelectrolyte content at a fixed NaCl concentration, or vice versa. In what follows the polyelectrolyte/lipid charge ratio $\xi$ is defined in terms of moles of the \emph{charged} lipid only, i.e. as the ratio $\xi = N_P/N_{L^+}$ between the nominal (i.e. assuming complete dissociation of the ionizable groups) number of moles of charges on the polyelectrolyte, $N_P$, and the number of moles of the charged lipid (DOTAP), $N_{L^+}$, in the liposomes.\\
The ionic strength of the aqueous medium was varied from 0.005 to 0.8 M by adding appropriate amounts of NaCl to the polyelectrolyte solution to be mixed with the liposome suspension. Since the contribution of the liposome and
polyelectrolyte counterions to the ionic strength is of the order of 1$-$2 mM, the largest contribution to the screening, except for the very low salt concentrations, comes from the added electrolyte.\\
A preliminary study was performed in order to check the stability of the different DOTAP/DOPC liposome suspensions in the presence of increasing amounts of added salt (but in absence of polyelectrolyte). Within the investigated NaCl concentration range (i.e. up to 0.8 M) the DOTAP/DOPC liposome suspensions are stable, the vesicles showing only a progressive reduction of their size (about 20$\%$), probably due to osmotic shrinkage \cite{SabinEPJE}, as it has been already observed for pure DOTAP liposomes \cite{Sennato04p296}. The measured $\zeta$ potential decreases at increasing the ionic strength, in agreement with the Gouy–Chapman theory \cite{Cevc93}, and consistently with the results of Barenholz et al. \cite{Hirsch-Lerner05} for similar DOTAP/DOPC liposome suspensions.

\begin{figure}[!ht]
    \includegraphics[width=8 cm]{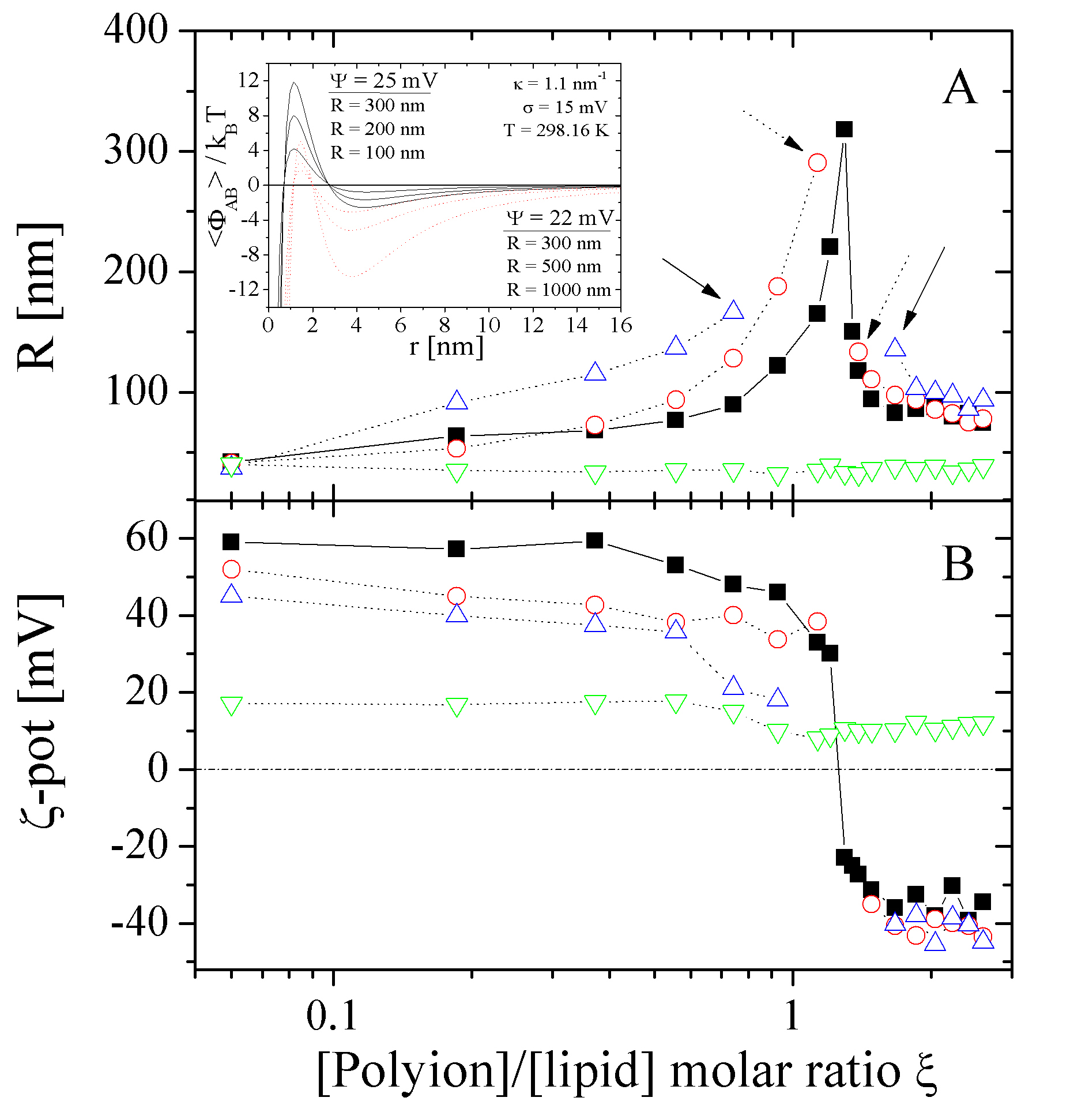}\\
  \caption{(Color online) The average hydrodynamic radius (panel A) and the corresponding $\zeta$-potential (panel B) of pd-liposome clusters formed by DOTAP/DOPC  0.5/0.5 liposomes as a function of the polyion/lipid molar ratio $\xi$, without added salt ($\blacksquare$) and in the presence of different concentrations of NaCl: 0.05 M ($\bigcirc$); 0.1 M ($\bigtriangleup$); 0.5 M ($\bigtriangledown$). The temperature was fixed at 25 $^\circ$C.  Panel A, inset: the potential of mean force $\langle\Phi_{AB}(r)\rangle$ of equation (\ref{eq:potential}) calculated for several values of the aggregate radii and at two different values of the surface electrostatic potential $\Psi$ : 25 mV (full lines) and 22 mV ; as in figure (\ref{Fig:Potential}) $\sigma_A=\sigma_B= 15\, mV$, $T= 298.16 \, K^\circ$, and $\mathcal{H} = 10^{-20} \, J $.  The inverse Debye constant is $\kappa= 1.1\, nm^{-1}$. At this ionic strength a small difference in the surface potential moves the system from the "barrier-limited"  to the "secondary-minimum" aggregation regime. }\label{Fig:PZ05}
\end{figure}

In Fig. \ref{Fig:PZ05} we show an example of the behavior of the mixed DOTAP/DOPC pd-liposomes in the $R-\xi$ and $\zeta-\xi$ planes. The curves are obtained for liposomes prepared at equimolar concentrations of DOTAP and DOPC at different salt content and are compared with the curves obtained for the pd-liposomes without any added salt.\\
As expected on the basis of the universality of this kind of aggregation, that does not depend on the details of the chemical structure of the polyelectrolyte and the colloid, in the absence of salt we found the typical reentrant condensation curve: stable pd-liposomes aggregates form in the whole range of $\xi$, except close to the isoelectric point. The addition of 0.05 M NaCl already causes a significant widening of the instability gap (marked by the dashed arrows in panel A of figure \ref{Fig:PZDOTAP}, and the addition of increasing amounts of salt makes the gap progressively larger. Notably, after 0.1 M NaCl, considering the same amount of added salt, the instability region widens at a lower charge ratio if compared with the one observed with pure DOTAP liposomes. That is, at the same polyion/lipid ratio $\xi$, pd-liposome aggregates formed by DOTAP-DOPC liposomes are destabilized by a lower amount of salt compared to DOTAP liposomes.
This finding is consistent with the expectations based on the potential described by equation (\ref{eq:potential}). Indeed, it is easy to calculate that the amount of added salt needed to move from the "barrier-limited" aggregation regime, where the size of the stable aggregates is determined by the increase of the potential barrier with the size of the aggregates, to the "secondary-minimum" regime, where the deepening of the secondary minimum causes the instability, decreases with the value of the surface potential (see the inset of figure (\ref{Fig:PZ05} - panel A). In other words, the secondary minimum deepens comparatively faster with the increasing salt content for lower values of the surface electrostatic potential $\Psi$.\\
\begin{figure}[!ht]
  \includegraphics[width=8cm]{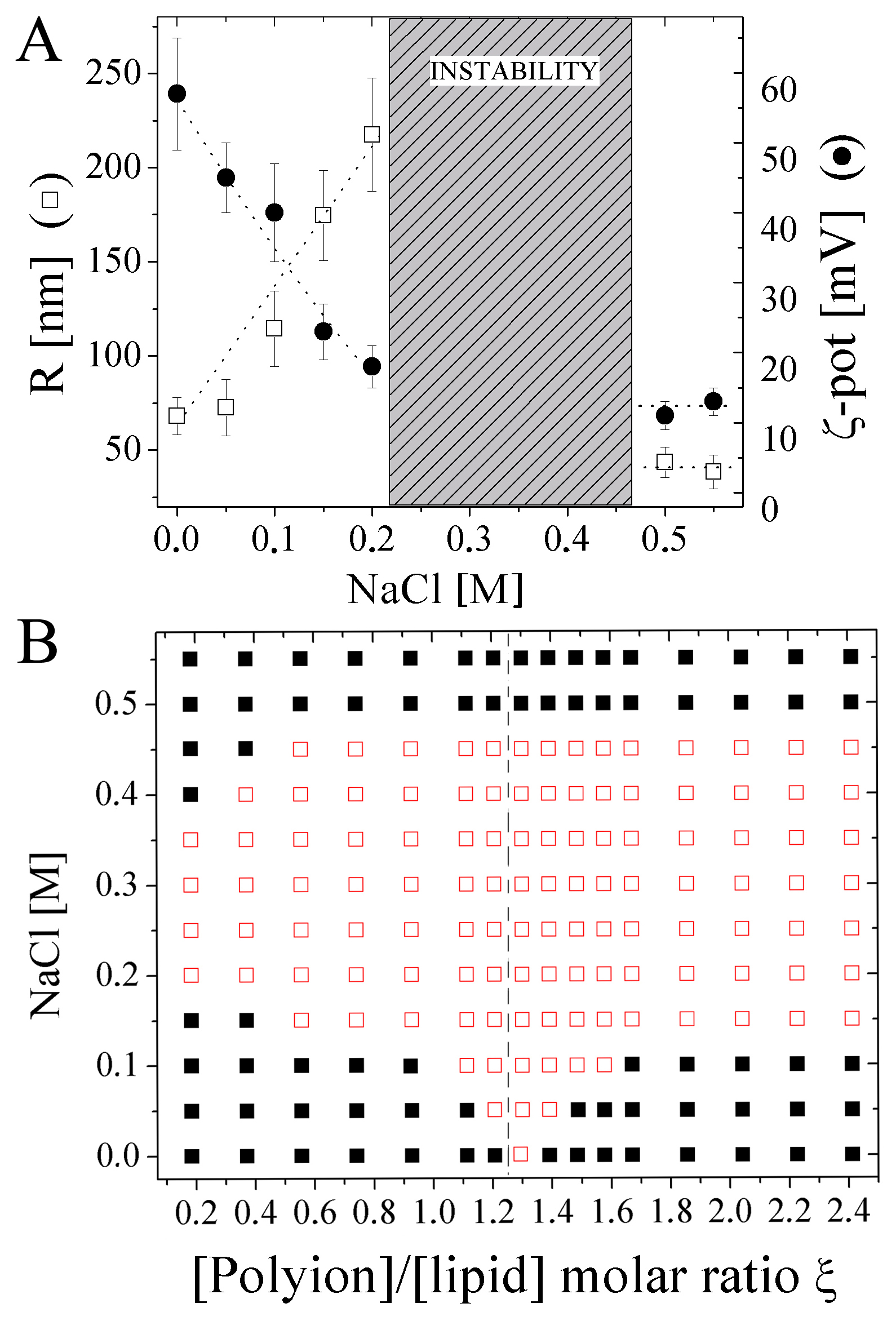}\\
  \caption{Panel A: the average hydrodynamic radius ($\square$) and the corresponding $\zeta$-potential ($\bullet$) of pd-liposome clusters formed by mixed DOTAP/DOPC  0.5/0.5  liposomes as a function of the added NaCl at a polyion/lipid molar ratio $\xi$= 0.4. Panel B: phase diagram of the DOTAP/DOPC  0.5/0.5 pd-liposome complexes formed in the presence of added salt. At each given polyion/lipid molar ratio $\xi$ and NaCl concentration, full symbols represent stable samples, open symbols represent the unstable ones. The dashed line marks the position of the isoelectric point, where the inversion of $\zeta$-potential occurs. As in all the previous experiments the temperature was fixed at T= 25 $^{\circ}$C.}\label{Fig:RZxi04}
\end{figure}
The addition of 0.15 M NaCl it is sufficient to destabilize the pd-liposomes aggregates at all the charge ratios larger than $\xi > 0.4$. Conversely, at  $\xi < 0.4$ small stable complexes are observed, with a radius of $100\div 200$ nm and a $\zeta$-potential of $\backsim 20 mV$. These values are significantly different from those characterizing the bare liposomes (R $\backsim$ 40 nm, $\zeta$ pot $\backsim 40$ mV), thus indicating that these particles are small aggregates of pd-liposomes. For a larger amount of salt, up to 0.4 M NaCl, a complete salt-induced destabilization of the pd-liposome is observed, at all the charge ratio $\xi$. Unexpectedly, a further salt addition promotes a return to small stable complexes. These complexes are observed only at low $\xi$ values in the presence of 0.45 M NaCl and over the whole $\xi$ range explored at 0.5 M, as it is shown in fig. \ref{Fig:PZDOTAP} ($\bigtriangledown$ symbols). The fact that the radius of these objects is around 50 nm, close the one measured for pure liposomes with same NaCl concentration added in solution, suggest that they are bare liposomes with scarce or null polyion adsorption.\\
Panel B of fig. (\ref{Fig:PZ05}) shows the corresponding behavior of the $\zeta$-potential of the samples whose data are shown in panel A. Again, the curve obtained in absence of added salt is reported (black circles) for comparison. As observed for pure DOTAP liposomes, shown in fig. (\ref{Fig:PZDOTAP}), the addition of salt up to 0.15 M promotes an evident decrease of the $\zeta$-potential at $\xi<1$, due to a progressive screening effect. In the overcharged region the same amount of added salt does not cause significant differences in the $\zeta$-potential, which remains approximately constant.\\
Above 0.4 M NaCl, a different behavior is found, which is not observed in pure DOTAP liposomes. What is new here is that the charge inversion is not observed anymore, but the $\zeta$-potential stays positive over the whole $\xi$ range investigated, with only a weak decrease above $\xi$=1. Notably, as for the hydrodynamic radius (panel A), the value of the $\zeta$-potential of DOTAP/DOPC 0.5/0.5 liposomes in the presence of 0.5 M NaCl does not change appreciably with or without the polyelectrolyte. Actually, above 0.4 M NaCl, both the size and $\zeta$-potential of the pd-liposomes correspond to those of the bare liposomes at the same concentration of salt, a clear indication that in this condition, due to the screening effect of the electrolyte the polyelectrolyte adsorption on the particle surface is very much reduced or absent.\\
In order to gain further insight on the effect on the aggregation induced by the added salt, it is instructive to considered the behavior of size and $\zeta$-potential of the aggregates as a function of the salt concentration and at a fixed polyion/lipid charge ratio.  An example is shown in panel A of Fig. (\ref{Fig:RZxi04})  for the complexes of DOTAP/DOPC  0.5/0.5 liposomes at $\xi$= 0.4.  A three phase behavior appears, ruled by salt concentration. At low salt, an increase of the NaCl content promotes the formation of stable pd-liposomes aggregates with increasing size and decreasing $\zeta$-potential, until the "secondary-minimum driven" instability region is reached. Beyond a critical concentration $C^*$, approximately between 0.45 and 0.5 M, a new region of coexisting "free" liposomes and polyelectrolytes is found. The polyelectrolyte does not adsorb anymore on the liposome.\\
Panel B of Fig. (\ref{Fig:RZxi04}) shows this behavior in terms of a phase diagram, analogous to that shown in figure (\ref{fig:DFDOTAP}) for pure DOTAP pd-liposomes. In this diagram the onset of a stable region of isolated bare liposomes at high salt content is evident, in contrast with the case of pure DOTAP pd-liposomes (fig. (\ref{fig:DFDOTAP})). Clearly, the aggregation of pd-liposomes can be enhanced or suppressed in a non trivial way by changing the ionic strength of the solution.\\
The behavior described for mixed DOTAP/DOPC 0.5/0.5 pd-liposomes has been observed with similar features in all the other mixed liposome preparations.\\
\begin{figure}[!ht]
  \includegraphics[width=5cm]{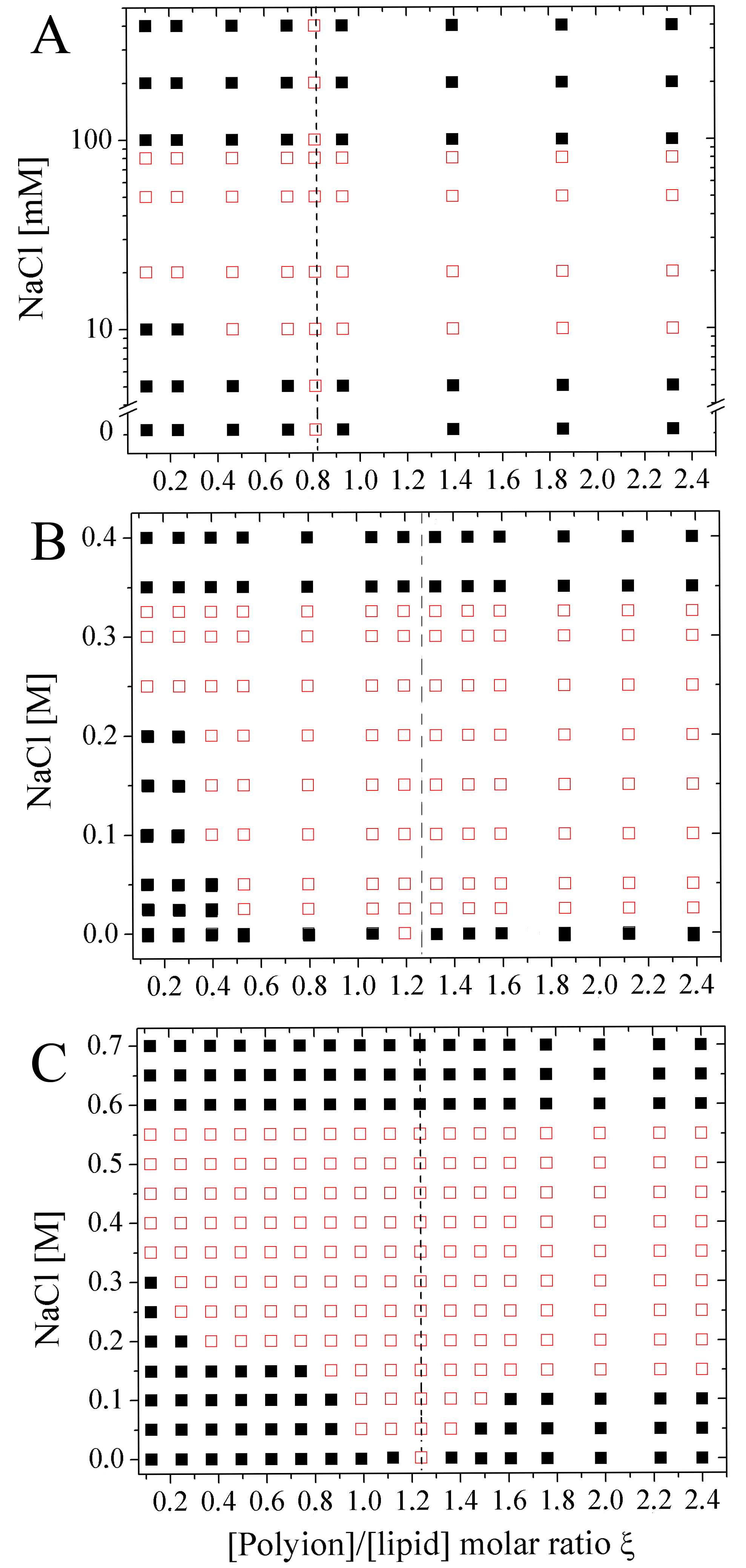}\\
  \caption{(Color online) Phase diagrams of the pd-liposome complexes formed by mixed DOTAP/DOPC liposomes in the presence of added salt, at the different DOTAP molar fractions (from top to bottom) X=0.2 (A), X = 0.35 (B) X = 0.75 (C). At each given polyion/lipid molar ratio $\xi$ and NaCl concentration, full symbols represent stable samples, open symbols represent the unstable ones. Dashed lines mark the position of the isoelectric point, where the inversion of $\zeta$-potential occurs. As in all the previous experiments the temperature was fixed at T= 25 $^{\circ}$C.}  \label{fig:DfaseMix}
\end{figure}
Fig. (\ref{fig:DfaseMix}) shows the phase diagrams of pd-liposomes built up at different DOTAP molar fractions X (with X increasing from top to bottom: panel A, X=0.2; B , X = 0.35; C, X = 0.75). For all the liposome preparations, a complete series of experiments was performed, up to 0.5 M NaCl at least, and in many cases up to 1 M NaCl. Differently from the case of pure DOTAP pd-liposomes (fig. (\ref{fig:DFDOTAP})), all the phase diagrams of mixed DOTAP-DOPC pd-liposomes are characterized by the presence of a stable region at high salt content. Within this region, independently of the charge ratio, stable complexes are observed with a size which remains close to that of the primary liposomes. The onset of this region for the different DOTAP/DOPC mixed liposomes occurs at different salt concentration. Such critical salt content increases when the molar fraction X of the charged lipid DOTAP composing the liposomes is increased.\\

Notably, for all the mixed DOTAP-DOPC liposomes with a DOTAP molar fraction larger than X=0.3 (including the pure DOTAP liposomes, fig (\ref{fig:DFDOTAP})), the isolectric point occurs at $\xi > 1$. Only for the liposome suspension with the lowest DOTAP concentration (X = 0.2) the charge inversion is observed at a charge ratio $< 1$, i.e. at $\xi \approx 0.8$. With the given definition of the stoichiometric polyelectrolyte/lipid charge ratio $\xi = N_P/N_{L+}$, in which the presence of the neutral DOPC lipid is not considered, a fixed position of the isoelectric point would imply a regular "compensation" between the decreased liposome charge and the amount of polyelectrolyte which is necessary to neutralize it.\\
In several studies on complexes formed by DNA and equimolar DOTAP-DOPC or DOTAP-DOPE mixed liposomes, it was observed that the charge neutralization point occurs at the same charge ratio of pure DOTAP liposomes \cite{zuidam1999characterization, Safinya99}.
Among these investigations, Koltover et al. \cite{Safinya99} varied the fraction of DOPC in mixed liposomes from 0 to 0.7, studying the structure of the resulting aggregates (lipoplexes) in pure water and in the presence of 0.15 M of NaCl. For DOTAP fractions smaller than 0.25, in excess of the neutral DOPC lipid, they observed a lipid demixing when DNA adsorbs to the liposome surface. Interestingly, DOTAP-DOPC liposomes also demix in the absence of DNA at the same DOPC fraction, forming stable lipid bilayers with the DOPC excess segregating in one-phase domains. A similar effect has been also observed in mixed DOTAP/DPPC monolayers \cite{Bordi08} and liposomes \cite{Cinelli07}, where in absence of polyelectrolytes at room temperature and at low molar fractions of DOTAP ($X \approx 0.2$) the mixtures show a tendency to separate in DOTAP-rich and DOTAP-poor domains.\\
As consequence of this demixing, in 0.2-0.8 DOTAP-DOPC liposomes an heterogeneous charge distribution may occur due to the formation of one-phase uncharged DOPC regions and mixed charged DOTAP-DOPC region. Spontaneous  "charge patches" can form at the bare liposome surface, thus giving rise to an attractive contribution which adds to the "charge patch" attraction due to correlated adsorption.  So in the 0.2-0.8 DOTAP-DOPC mixed liposomes the (total) attraction is probably larger compared the other mixed liposomes, at the same polyion coverage, i.e, at the same $\xi$, and a smaller amount of polyelectrolyte is sufficient to reduce the repulsions via charge neutralization and to promote the aggregation of like-charged pd-liposomes. Such demixing could justify the occurrence of the isoelectric point and of the aggregation maximum at a lower value of $\xi$ for the X=0.2 mixture (fig. (\ref{fig:DfaseMix}) panel A). Indeed, a smaller number of adsorbed chains with loops and dangling ends can efficiently neutralize the charged lipids when they are concentrated in a few domains (more chains would be needed where the charged lipids distributed more uniformly).\\

The possibility to displace the adsorbed polyions from the surface of liposomes upon increasing the concentration of salt in the solution has been observed also in other liposome systems. Yaroslavov and coworkers  \cite{Yaroslavov11} observed the desorption of the positively charged polycation (poly-N-ethyl-4-vinylpyridinium bromide) from the surface of anionic liposomes (composed of phosphatidylserine (PS) and phosphatidylcholine (PC)) when a critical concentration of salt is reached.  Since  the  salt concentration needed to completely displace the polyelectrolyte from the surface is roughly proportional to the amount of the charged lipid (PS) in the liposome composition, the authors have hypothesized that the desorption is basically controlled by the surface charge density. In that work the liposomes are formed by two naturally occurring mixtures of PS and PC lipids. In this system, since the exact composition of the hydrophobic tails of the lipid is unknown, the determination of the area per molecule occupied at the bilayer surface, and hence the surface charge density of the liposomes, is uncertain. So that it is comparatively more difficult to correlate the effect of the salt and of the surface charge on the adsorption/desorption of the polyelectrolyte.\\

\begin{figure}[!ht]
  \includegraphics[width=8cm]{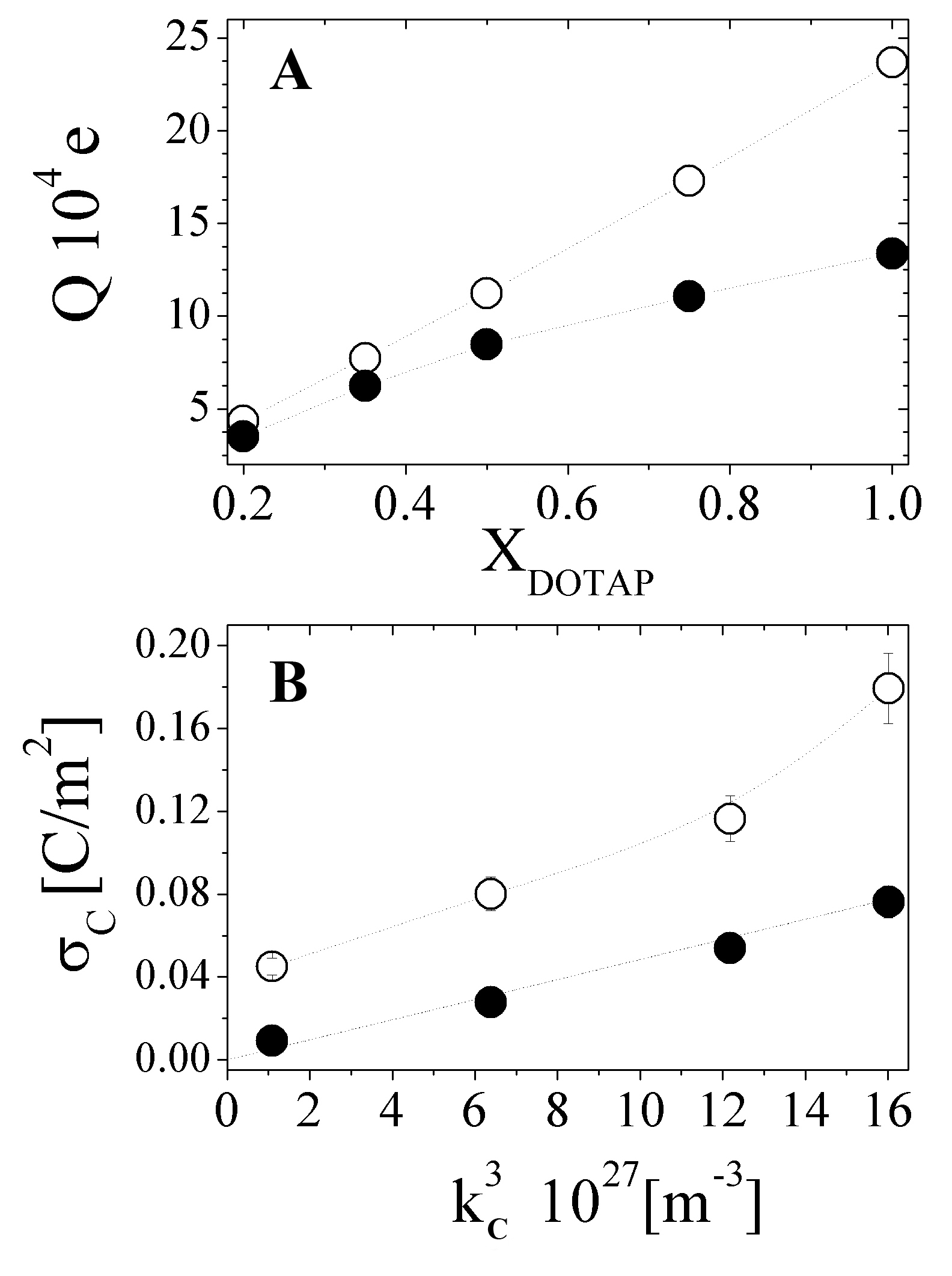}\\
  \caption{Panel A, the nominal ($\circ$) and effective ($\bullet$) charge, calculated according to the Aubouy's model, as a function of the molar fraction of the cationic lipid in the mixed liposomes, lines guides the eye only. Panel B, when plotted against the cube of the experimental inverse Debye length of desorption, $\kappa_c$, the nominal surface charge of the DOTAP/DOPC mixed liposomes ($\circ$) does not show the linear trend through the origin expected from equation (\ref{eq:winkler}) (the line is a guide for the eye only). On the contrary, when the renormalized charge calculated from Aubouy's model (\ref{eq:Auboy1}) is employed the agreement is excellent (($\bullet$), coefficient of determination, R-squared= 0.949, P$< 0.026$).}\label{FigSoglie}
\end{figure}

These findings can be rationalized within the theory first developed by Muthukumar \cite{Muthukumar87} for pure electrostatic adsorption, proving that polyion adsorption can be finely controlled by changing the surface charge of the liposomes on the molecular scale and, more generally, by tuning the electrostatic interaction, which, in turn, rule the stability of the pd-liposome complexes.\\
The effect on the adsorption/desorption of a polyelectrolyte onto an oppositely charged surface of the presence of added salt in the aqueous solution bathing the surface has been investigated theoretically by several authors \cite{Wiegel77,Muthukumar87,vonGoeler94,Haronska98,Hansupalak03,Winkler06}. In particular, based on a generalized mean-field approach, Winkler and Cherstvy \cite{Winkler06} proposed a model for the adsorption of a flexible polyelectrolyte on a spherical surface. According to this model, the non-adsorption regime occurs when the inverse of the Debye screening length $\kappa$ exceeds a critical value:
\begin{equation}\label{eq:winkler}
    \kappa_c^3 = \frac{24|\sigma_c||\rho|}{j^2_0 \varepsilon_0 \varepsilon_r k_B T l_K}
\end{equation}
where $ j_0= 2.405$ is the first positive root of the Bessel functions of the first kind $J_0$, $l_K$ is the Kuhn length of the polyelectrolyte, $\sigma_c$  and $\rho$ the surface and linear charge density of spherical surface and polyelectrolyte, respectively. Equation (\ref{eq:winkler}) holds in the limit $\kappa R \rightarrow \infty$, where $R$ is the particle radius. In our experiments $\kappa R$ is typically $\approx 100$.\\
From the experimental NaCl vs $\xi$ phase diagram of the mixed DOTAP-DOPC pd-liposomes (figure (\ref{Fig:RZxi04} panel B and (\ref{fig:DfaseMix})) the critical concentration of NaCl needed to induce the polyion desorption can be estimated as midway between the last line of "unstable" samples and the the first of stable ones. From these values, assuming that the contribution of the counterions of the liposomes and the polyelectrolyte is negligible, a critical screening length $\kappa_c^{-1}$ can be determined. In this way, since the structural parameters that appear in equation (\ref{eq:winkler}), the Kuhn length etc., of our system are known from the literature, this equation can be used to analyze the observed dependence of the critical salt concentration for desorption on the liposome surface charge.\\
The persistence length (which is half the Kuhn length) of NaPA is $\approx 12 \, ${\AA} \cite{Zhang00}, so that $l_K \approx 24 \, ${\AA}. The distance between the (monovalent) ionizable groups of NaPA is $b = 2.52 ${\AA} \cite{Mylonas99}, so that the nominal (i.e. without taking into account the counterion condensation) linear charge density on the chain can be readily determined.\\
The stoichiometric surface charge density of mixed DOTAP-DOPC liposomes can be estimated considering the values of the area per molecule occupied in a bilayer by pure DOPC and DOTAP, 72.5 and 70  {\AA}$^2$ respectively \cite{Petrache04}. Then, assuming the cationic DOTAP headgroup completely ionized and the DOPC (which is zwitterionic) completely neutral, the surface charge density  $\sigma_c$ can be calculated as a function of the composition as
\begin{equation}\label{eq:sigma}
    \sigma_c= \frac{e}{[A_{DOTAP}+A_{DOPC}(\frac{1}{X}-1)]}
\end{equation}
where $e$ is the unit charge, $X$  is the DOTAP molar fraction in the mixture, and $A_{DOTAP}$ and $A_{DOPC}$ are the values of the area per molecule of DOTAP and DOPC, respectively. Actually, however, the nominal or "bare" charge, $Q_{bare}$, of these liposomes must be replaced by a renormalized one \cite{Quesada-Perez02}, since, due to the rather high density of ionizable groups on the surface, part of the counterions remain "entrapped" in its immediate vicinity. Indeed, experimentally, the effective charge $Q_{eff}$ of many colloids is considerably smaller than the "structural" surface charge estimated, for example, by titration \cite{haro2003}. The relation between bare charge and effective charge of mixtures of charged and zwitteionic lipids has been thoroughly investigated for the DOTAP-DOPE system \cite{Bordi06e}, and both in the case of spherical vesicles (liposomes), and of planar monolayers (Langmuir monolayers) it has been shown that the effective charge can be rather satisfactorily evaluated applying the model of Aubouy et al. \cite{Auboy03}.\\
Based on an approximate nonlinear Poisson-Boltzmann equation, in the limit of large $k_D R$ the model gives an equation relating the effective charge $Q_{eff}$ and the bare charge $Q_{bare}= eZ_{bare}$ of spherical colloids
\begin{equation} \label{eq:Auboy1}
    Q_{eff} \frac{l_B}{R}= 4k_D R T(x)+ 2 \left[  5-\frac{T(x)^4 +3}{T(x)^2 +1} \right] T(x)
\end{equation}
where the function $T(x)$ is defined as
\begin{equation}
    T(x)= \frac{1}{x} (\sqrt{1+x^2}-1)
\end{equation}
and $x$ as
\begin{equation}
    x= \frac{Z_{bare}l_B}{R(2k_D R+2)}
\end{equation}
Equation (\ref{eq:Auboy1}) can be employed to calculate the effective surface charge density of mixed DOTAP-DOPC liposomes at the NaCl concentration corresponding to the desorption condition that must be used in eq. (\ref{eq:winkler}) to fit the experimental data. Panel A of fig. (\ref{FigSoglie}) shows the behavior of the nominal and the effective liposome charge (full and open symbols, respectively) calculated for the liposomes prepared at the different molar fractions of the charged lipid (DOTAP) in the mixture. As expected, at increasing liposome surface charge the renormalization becomes more relevant.\\

Also in the case of the polyelectrolyte employed, NaPA, the linear charge density should be reduced by a factor $f$, due to the counterion condensation phenomena. For polyelectrolytes in the infinite dilution limit, Manning's rule allows to calculate $f$ as $f=b/l_B$, where $l_B$ is the Bjerrum length  $e^2/(4\pi \varepsilon_0 \varepsilon_r k_BT)$, about 7.14 {\AA} in water at T=25 $^\circ$C,  which gives for NaPA  $f=0.35$ \cite{Bordi04b}. However, experimentally both the persistence length and the fraction of uncondensed counterions changes with the concentration and with the ionic strength of the solution (for $f$, see for example \cite{Bordi02}).  Moreover, the  determination of  the effective polyion charge in adsorption condition is a controversial issue. Different theories show that in the process of adsorption of polyelectrolytes onto an oppositely charged colloidal particle the counterions can be completely released \cite{fleck2001counterion}, mostly released \cite{park1999spontaneous},
partially released \cite{nguyen2000macroions}, not always released \cite{sens2000counterion}, or not released at all (in the case of flexible polyelectrolytes) \cite{dobrynin2000adsorption}.\\
As a check of the validity of the model represented by equation (\ref{eq:winkler}) to describe the polyelectrolyte desorption in our system, we plotted the surface charge densities of the DOTAP-DOPC liposomes prepared at the different concentrations of the charged lipid versus the cube of the  "critical" inverse Debye length, $\kappa_c$, obtained experimentally from the phase diagrams of figs. (\ref{Fig:RZxi04})- Panel B and (\ref{fig:DfaseMix}). If equation (\ref{eq:winkler}) holds a linear trend passing through the origin of the $\sigma_C-\kappa_C^3$ plane is expected.\\
The result is shown in panel B of fig. (\ref{FigSoglie}). The open symbols represent the nominal values of the surface charge densities calculated as a function of the charged lipid molar fraction from equation (\ref{eq:sigma}). In this case the relation (\ref{eq:winkler}) is not satisfied. However, if the renormalized values calculated from the Aubouy's model (\ref{eq:Auboy1}) are used (full symbols) the agreement is excellent (R-squared value 0.949, P $<$ 0.026). From the value of the slope obtained from the linear fit, being all the other quantities known, the factor $f$, introduced in order to renormalize the polyion charge density $\rho$ can be calculated. It is worth to remind here that this is the effective charge of the polyelectrolyte when it is \emph{adsorbed} on the oppositely charged particle surface. In practice $f$ can be interpreted as the fraction of the counterions that are not condensed, i.e. do not keep staying in the proximity of the chain, but are "free" in the solution, when the polyion is adsorbed. The value of $f$ obtained from the fit is $f=0.55 \pm 0.1$, thus indicating that the counterions are only \emph{partially released} in the process of adsorption. This value for the fraction of free counterions can be compared with the analogous value for the polyelectrolyte when it is not adsorbed. As we stated above, the Manning's rule gives for NaPA a values $f=0.35$. Experimentally, for this highly charged polyelectrolyte a value of $f=0.25\div 0.30$ has been estimated \cite{Bordi02} from low frequency conductivity measurements, in aqueous solution at T=25 $^\circ$C and in absence of added salt. This value increases up to $f=0.4\div0.6$ in the presence of a large excess of added salt \cite{Bordi03pre}.\\
In addition to that, a partial release of the condensed counterions when the polyelectrolyte adsorbs on an oppositely charged particle has been also observed in the case of NaPA adsorbed on pure DOTAP liposomes, with a completely different experimental approach based on electrical conductivity measurements \cite{Bordi06}.

\section{Conclusions}
We have investigated the "reentrant condensation" and the charge inversion of polyelectrolyte-decorated liposomes in a wide range of surface charge densities of  liposomes, from $\approx 0.25 $ C/m$^2$ of pure DOTAP liposomes \cite{SennatoSM2012} to $\approx 0.05$ C/m$^2$, correspondent to the DOTAP/DOPC mixture with only $20\%$ DOTAP.\\
In all cases the Velegol-Twar potential \cite{Velegol01}, extended by the inclusion of the van der Waals interactions \cite{Truzzolillo09,Truzzolillo10}, well captures the observed phenomenology.\\
When the surface charge density of the mixed liposomes is lowered via progressively adding neutral lipid within the liposome membrane, a new intriguing phenomenon shows up when the concentration of a simple electrolyte in the dispersing medium increases above a critical value: the stability diagram of the suspensions shows a novel reentrance due to the crossing of the desorption threshold of the polyelectrolyte. Indeed at fixed charge density of the bare DOTAP/DOPC liposomes and for a wide range of polyion concentrations, we showed that the simple electrolyte addition first (low salt regime) destabilizes the suspensions because of the enhanced screening of the residual repulsion between the complexes, and then (high salt regime) determines the onset of a new stable phase originated by the absence of polyelectrolyte adsorption on the particle surfaces.
At a fixed linear charge density of the polyelectrolyte, the onset of this desorption regime depends on both the ionic strength of the bathing solution and on the liposome surface charge density. Once an effective surface charge density, taking into account the effect of the counterions, is calculated using the Aubuoy's model \cite{Auboy03} for each DOTAP/DOPC mixture, the dependence of this effective charge from the cube of the critical inverse Debye length $\kappa_c$ shows an excellent agreement with the linear trend predicted by the adsorption theory of Winkler and Cherstvy \cite{Winkler06}. Via a linear fit we estimated a value of the renormalized linear charge density on the polyelectrolyte which is consistent with previous estimates based on independent measurements \cite{Bordi06}.\\
These findings give strong evidence that, in presence of adsorption from a bathing aqueous solution of highly charged polyelectrolytes onto an oppositely charged colloids, both the adsorbing chains and the particles maintain part of their "condensed" counterions, and give further support to the overall picture of the reentrant condensation of pd-colloids as driven by the correlated adsorption of the polyelectrolyte on the particle surface.\\
In the wider framework of bionanotechnology, this investigation find its natural applications in the engineering of nanoparticle systems, having implications in the definition of its colloidal stability and also of its biological fate, which it has been shown to be strongly influenced by the formation of the so called "protein corona" \cite{monopoli2011physical}. Actually, these two points are closely related each other, since the protein corona itself contributes to shape the surface properties, the charge and the resistance to aggregation of the nanoparticles. It was also shown that the adsorption of proteins onto the nanoparticle surfaces can strongly affect the protein conformation, which may ultimately culminate in the loss of its biological activity \cite{wang2011soft}. In this view, these results give further support to the possibility to trigger the protein adsorption at the surface of charged nanoparticles in a physiological environment by tuning the surface charge. This last aspect is also envisaged as relevant for the perspective to include strategically a protein corona into designed composite nanoparticles for the load and the controlled release of different drugs \cite{kah2012exploiting}.

\newpage
\bibliography{BiblioArt}

\begin{thebibliography}{81}
\expandafter\ifx\csname natexlab\endcsname\relax\def\natexlab#1{#1}\fi
\providecommand{\url}[1]{\texttt{#1}}
\providecommand{\href}[2]{#2}
\providecommand{\path}[1]{#1}
\providecommand{\DOIprefix}{doi:}
\providecommand{\ArXivprefix}{arXiv:}
\providecommand{\URLprefix}{URL: }
\providecommand{\Pubmedprefix}{pmid:}
\providecommand{\doi}[1]{\href{http://dx.doi.org/#1}{\path{#1}}}
\providecommand{\Pubmed}[1]{\href{pmid:#1}{\path{#1}}}
\providecommand{\bibinfo}[2]{#2}
\ifx\xfnm\relax \def\xfnm[#1]{\unskip,\space#1}\fi
\bibitem[{Winkler and Cherstvy(2006)}]{Winkler06}
\bibinfo{author}{R.~G. Winkler}, \bibinfo{author}{A.~G. Cherstvy},
\newblock \bibinfo{title}{Critical adsorption of polyelectrolytes onto charged
  spherical colloids},
\newblock \bibinfo{journal}{Phys. Rev. Lett.} \bibinfo{volume}{96}
  (\bibinfo{year}{2006}) \bibinfo{pages}{066103--6}.
\bibitem[{Volkenstein(1977)}]{Volkenstein}
\bibinfo{author}{M.~Volkenstein}, \bibinfo{title}{Molecular Biophysics},
  \bibinfo{publisher}{Academic Press, New York}, \bibinfo{year}{1977}.
\bibitem[{Napper(1983)}]{Napper}
\bibinfo{author}{D.~H. Napper}, \bibinfo{title}{Polymeric Stabilization of
  Colloidal Dispersions}, \bibinfo{publisher}{Academic Press, New York},
  \bibinfo{year}{1983}.
\bibitem[{Amenitsch et~al.(2011)Amenitsch, Caracciolo, Foglia, Fuscoletti,
  Giansanti, Marianecci, Pozzi, and Lagan{\`a}}]{amenitsch2011existence}
\bibinfo{author}{H.~Amenitsch}, \bibinfo{author}{G.~Caracciolo},
  \bibinfo{author}{P.~Foglia}, \bibinfo{author}{V.~Fuscoletti},
  \bibinfo{author}{P.~Giansanti}, \bibinfo{author}{C.~Marianecci},
  \bibinfo{author}{D.~Pozzi}, \bibinfo{author}{A.~Lagan{\`a}},
\newblock \bibinfo{title}{Existence of hybrid structures in cationic
  liposome/\text{DNA} complexes revealed by their interaction with plasma
  proteins},
\newblock \bibinfo{journal}{Coll. Surf. B} \bibinfo{volume}{82}
  (\bibinfo{year}{2011}) \bibinfo{pages}{141--146}.
\bibitem[{Levin(2002)}]{levin2002electrostatic}
\bibinfo{author}{Y.~Levin},
\newblock \bibinfo{title}{Electrostatic correlations: from plasma to biology},
\newblock \bibinfo{journal}{Rep. Prog. Phys.} \bibinfo{volume}{65}
  (\bibinfo{year}{2002}) \bibinfo{pages}{1577}.
\bibitem[{Szilagyi et~al.(2014)Szilagyi, Trefalt, Tiraferri, Maroni, and
  Borkovec}]{Borkovec14p03}
\bibinfo{author}{I.~Szilagyi}, \bibinfo{author}{G.~Trefalt},
  \bibinfo{author}{A.~Tiraferri}, \bibinfo{author}{P.~Maroni},
  \bibinfo{author}{M.~Borkovec},
\newblock \bibinfo{title}{Polyelectrolyte adsorption, interparticle forces, and
  colloidal aggregation},
\newblock \bibinfo{journal}{Anal. Chem.} \bibinfo{volume}{46}
  (\bibinfo{year}{2014}) \bibinfo{pages}{3}.
\bibitem[{Nguyen et~al.(2000)Nguyen, Grosberg, and
  Shklovskii}]{nguyen2000macroions}
\bibinfo{author}{T.~T. Nguyen}, \bibinfo{author}{A.~Y. Grosberg},
  \bibinfo{author}{B.~I. Shklovskii},
\newblock \bibinfo{title}{Macroions in salty water with multivalent ions: giant
  inversion of charge},
\newblock \bibinfo{journal}{Phys. Rev. Lett.} \bibinfo{volume}{85}
  (\bibinfo{year}{2000}) \bibinfo{pages}{1568}.
\bibitem[{Agrati et~al.(2011)Agrati, Marianecci, Sennato, Carafa, Bordoni,
  Cimini, Tempestilli, Pucillo, Turchi, Martini, Borioni, and Bordi}]{Agrati10}
\bibinfo{author}{C.~Agrati}, \bibinfo{author}{C.~Marianecci},
  \bibinfo{author}{S.~Sennato}, \bibinfo{author}{M.~Carafa},
  \bibinfo{author}{V.~Bordoni}, \bibinfo{author}{E.~Cimini},
  \bibinfo{author}{M.~Tempestilli}, \bibinfo{author}{L.~P. Pucillo},
  \bibinfo{author}{F.~Turchi}, \bibinfo{author}{F.~Martini},
  \bibinfo{author}{G.~Borioni}, \bibinfo{author}{F.~Bordi},
\newblock \bibinfo{title}{Multicompartment vectors as novel drug delivery
  systems: selective activation of \text{T}$\gamma\delta$ lymphocytes after
  zoledronic acid delivery},
\newblock \bibinfo{journal}{Nanomedicine} \bibinfo{volume}{7}
  (\bibinfo{year}{2011}) \bibinfo{pages}{153--161}.
\bibitem[{Bordi et~al.(2014)Bordi, Chronopoulou, Palocci, Bomboi, Martino,
  Cifani, Pompili, Ascenzioni, and Sennato}]{Bordi14p184}
\bibinfo{author}{F.~Bordi}, \bibinfo{author}{L.~Chronopoulou},
  \bibinfo{author}{C.~Palocci}, \bibinfo{author}{F.~Bomboi},
  \bibinfo{author}{A.~D. Martino}, \bibinfo{author}{N.~Cifani},
  \bibinfo{author}{B.~Pompili}, \bibinfo{author}{F.~Ascenzioni},
  \bibinfo{author}{S.~Sennato},
\newblock \bibinfo{title}{{Chitosan-\text{DNA} complexes: Effect of molecular
  parameters on the efficiency of delivery}},
\newblock \bibinfo{journal}{Coll. Surf. A} \bibinfo{volume}{460}
  (\bibinfo{year}{2014}) \bibinfo{pages}{184 -- 190}.
\bibitem[{Bordi et~al.(2009)Bordi, Sennato, and Truzzolillo}]{Bordi09Review}
\bibinfo{author}{F.~Bordi}, \bibinfo{author}{S.~Sennato},
  \bibinfo{author}{D.~Truzzolillo},
\newblock \bibinfo{title}{Polyelectrolyte-induced aggregation of liposomes: a
  new cluster phase with interesting applications},
\newblock \bibinfo{journal}{J. Phys.: Cond. Matt.} \bibinfo{volume}{21}
  (\bibinfo{year}{2009}) \bibinfo{pages}{203102 (26pp)}.
\bibitem[{Quemeneur et~al.(2010)Quemeneur, Rinaudo, Maret, and
  Pepin-Donat}]{Quemeneur10}
\bibinfo{author}{F.~Quemeneur}, \bibinfo{author}{M.~Rinaudo},
  \bibinfo{author}{G.~Maret}, \bibinfo{author}{B.~Pepin-Donat},
\newblock \bibinfo{title}{Decoration of lipid vesicles by polyelectrolytes:
  mechanism and structure},
\newblock \bibinfo{journal}{Soft Matter} \bibinfo{volume}{6}
  (\bibinfo{year}{2010}) \bibinfo{pages}{4471--4481}.
\bibitem[{Safinya(2001)}]{safinya2001structures}
\bibinfo{author}{C.~R. Safinya},
\newblock \bibinfo{title}{Structures of lipid-\text{DNA} complexes:
  supramolecular assembly and gene delivery},
\newblock \bibinfo{journal}{Curr. Opin. Struct. Biol.} \bibinfo{volume}{11}
  (\bibinfo{year}{2001}) \bibinfo{pages}{440--448}.
\bibitem[{Bordi et~al.(2004)Bordi, Cametti, Diociaiuti, Gaudino, Gili, and
  Sennato}]{Bordi04}
\bibinfo{author}{F.~Bordi}, \bibinfo{author}{C.~Cametti},
  \bibinfo{author}{M.~Diociaiuti}, \bibinfo{author}{D.~Gaudino},
  \bibinfo{author}{T.~Gili}, \bibinfo{author}{S.~Sennato},
\newblock \bibinfo{title}{Complexation of anionic polyelectrolytes with
  cationic liposomes: Evidence of reentrant condensation and lipoplex
  formation},
\newblock \bibinfo{journal}{Langmuir} \bibinfo{volume}{20}
  (\bibinfo{year}{2004}) \bibinfo{pages}{5214--5222}.
\bibitem[{Sennato et~al.(2005)Sennato, Bordi, Cametti, Diociaiuti, and
  Malaspina}]{Sennato05p11}
\bibinfo{author}{S.~Sennato}, \bibinfo{author}{F.~Bordi},
  \bibinfo{author}{C.~Cametti}, \bibinfo{author}{M.~Diociaiuti},
  \bibinfo{author}{P.~Malaspina},
\newblock \bibinfo{title}{Charge patch attraction and reentrant condensation in
  \text{DNA}-liposome complexes},
\newblock \bibinfo{journal}{Biochim. Biophys. Acta} \bibinfo{volume}{1714}
  (\bibinfo{year}{2005}) \bibinfo{pages}{11--24}.
\bibitem[{Sennato et~al.(2004)Sennato, Bordi, and Cametti}]{Sennato04p296}
\bibinfo{author}{S.~Sennato}, \bibinfo{author}{F.~Bordi},
  \bibinfo{author}{C.~Cametti},
\newblock \bibinfo{title}{Correlated adsorption of polyelectrolytes in the
  charge inversion of colloidal particles},
\newblock \bibinfo{journal}{Europhys. Lett.} \bibinfo{volume}{68}
  (\bibinfo{year}{2004}) \bibinfo{pages}{296--302}.
\bibitem[{Grosberg et~al.(2002)Grosberg, Nguyen, and Shklovskii}]{Grosberg02}
\bibinfo{author}{A.~Y. Grosberg}, \bibinfo{author}{T.~T. Nguyen},
  \bibinfo{author}{B.~I. Shklovskii},
\newblock \bibinfo{title}{Colloquium: the physics of charge inversion in
  chemical and biological systems},
\newblock \bibinfo{journal}{Rev. Mod. Phys.} \bibinfo{volume}{74}
  (\bibinfo{year}{2002}) \bibinfo{pages}{329--345}.
\bibitem[{Dobrynin et~al.(2000)Dobrynin, Deshkovski, and
  Rubinstein}]{Dobrynin00}
\bibinfo{author}{A.~V. Dobrynin}, \bibinfo{author}{A.~Deshkovski},
  \bibinfo{author}{M.~Rubinstein},
\newblock \bibinfo{title}{Adsorption of polyelectrolytes at an oppositely
  charged surface},
\newblock \bibinfo{journal}{Phys. Rev. Lett.} \bibinfo{volume}{84}
  (\bibinfo{year}{2000}) \bibinfo{pages}{3101--3104}.
\bibitem[{Truzzolillo et~al.(2010)Truzzolillo, Bordi, Sciortino, and
  Sennato.}]{Truzzolillo10}
\bibinfo{author}{D.~Truzzolillo}, \bibinfo{author}{F.~Bordi},
  \bibinfo{author}{F.~Sciortino}, \bibinfo{author}{S.~Sennato.},
\newblock \bibinfo{title}{Interaction between like-charged
  polyelectrolyte-colloid complexes in electrolyte solutions: A \text{M}onte
  \text{C}arlo simulation study in the \text{D}ebye-\text{H}\"{u}ckel
  approximation},
\newblock \bibinfo{journal}{J. Chem. Phys.} \bibinfo{volume}{113}
  (\bibinfo{year}{2010}) \bibinfo{pages}{024901--10}.
\bibitem[{Sch{\"o}nhoff(2003)}]{schonhoff2003layered}
\bibinfo{author}{M.~Sch{\"o}nhoff},
\newblock \bibinfo{title}{Layered polyelectrolyte complexes: physics of
  formation and molecular properties},
\newblock \bibinfo{journal}{J. Phys.: Cond. Matt.} \bibinfo{volume}{15}
  (\bibinfo{year}{2003}) \bibinfo{pages}{R1781}.
\bibitem[{Volodkin et~al.(2007)Volodkin, Ball, Schaaf, Voegel, and
  Mohwald}]{Volodkin07}
\bibinfo{author}{D.~Volodkin}, \bibinfo{author}{V.~Ball},
  \bibinfo{author}{P.~Schaaf}, \bibinfo{author}{J.~Voegel},
  \bibinfo{author}{H.~Mohwald},
\newblock \bibinfo{title}{Complexation of phosphocholine liposomes with
  polylysine. stabilization by surface coverage versus aggregation},
\newblock \bibinfo{journal}{Biochim. Biophys. Acta} \bibinfo{volume}{1768}
  (\bibinfo{year}{2007}) \bibinfo{pages}{280--290}.
\bibitem[{Miklavic et~al.(1994)Miklavic, Chan, White, and Healy}]{Miklavic94}
\bibinfo{author}{S.~J. Miklavic}, \bibinfo{author}{D.~Y.~C. Chan},
  \bibinfo{author}{L.~R. White}, \bibinfo{author}{T.~W. Healy},
\newblock \bibinfo{title}{Double layer forces between heterogeneous charged
  surfaces},
\newblock \bibinfo{journal}{J. Phys. Chem.} \bibinfo{volume}{98}
  (\bibinfo{year}{1994}) \bibinfo{pages}{9022--9032}.
\bibitem[{Khachatourian and Wistrom(1998)}]{Khachatourian98}
\bibinfo{author}{A.~V.~M. Khachatourian}, \bibinfo{author}{A.~O. Wistrom},
\newblock \bibinfo{title}{Electrostatic interaction force between planar
  surfaces due to 3-d distribution of sources of potential (charge)},
\newblock \bibinfo{journal}{J. Phys. Chem. B} \bibinfo{volume}{102}
  (\bibinfo{year}{1998}) \bibinfo{pages}{2483--2493}.
\bibitem[{Velegol and Thwar(2001)}]{Velegol01}
\bibinfo{author}{D.~Velegol}, \bibinfo{author}{P.~Thwar},
\newblock \bibinfo{title}{Analytical model for the effect of surface charge
  nonuniformity on colloidal interactions},
\newblock \bibinfo{journal}{Langmuir} \bibinfo{volume}{17}
  (\bibinfo{year}{2001}) \bibinfo{pages}{7687--7693}.
\bibitem[{Mukherjee(2004)}]{Mukherjee04}
\bibinfo{author}{A.~K. Mukherjee},
\newblock \bibinfo{title}{The attraction between like-charged macroions: the
  crucial roles of macroion geometry and charge distribution},
\newblock \bibinfo{journal}{J. Phys.: Cond. Matt.} \bibinfo{volume}{16}
  (\bibinfo{year}{2004}) \bibinfo{pages}{2907}.
\bibitem[{Sennato et~al.(2008)Sennato, Truzzolillo, Bordi, and
  Cametti}]{Sennato08}
\bibinfo{author}{S.~Sennato}, \bibinfo{author}{D.~Truzzolillo},
  \bibinfo{author}{F.~Bordi}, \bibinfo{author}{C.~Cametti},
\newblock \bibinfo{title}{Effect of temperature on the reentrant condensation
  in polyelectrolyte-liposome complexation},
\newblock \bibinfo{journal}{Langmuir} \bibinfo{volume}{24}
  (\bibinfo{year}{2008}) \bibinfo{pages}{12181--8}.
\bibitem[{Bordi et~al.(2005)Bordi, Cametti, and Sennato}]{Bordi05p134}
\bibinfo{author}{F.~Bordi}, \bibinfo{author}{C.~Cametti},
  \bibinfo{author}{S.~Sennato},
\newblock \bibinfo{title}{Polyions act as an electrostatic glue for mesoscopic
  particle aggregates},
\newblock \bibinfo{journal}{Chem. Phys. Lett.} \bibinfo{volume}{409}
  (\bibinfo{year}{2005}) \bibinfo{pages}{134 -- 138}.
\bibitem[{Keren et~al.(2002)Keren, Soen, Ben~Yoseph, Gilad, Braun, Sivan, and
  Talmon}]{Keren02}
\bibinfo{author}{K.~Keren}, \bibinfo{author}{Y.~Soen},
  \bibinfo{author}{G.~Ben~Yoseph}, \bibinfo{author}{R.~Gilad},
  \bibinfo{author}{E.~Braun}, \bibinfo{author}{U.~Sivan},
  \bibinfo{author}{Y.~Talmon},
\newblock \bibinfo{title}{Microscopics of complexation between long \text{DNA}
  molecules and positively charged colloids},
\newblock \bibinfo{journal}{Phys. Rev. Lett.} \bibinfo{volume}{89}
  (\bibinfo{year}{2002}) \bibinfo{pages}{881031--4}.
\bibitem[{Zuzzi et~al.(2008)Zuzzi, Cametti, and Onori}]{Zuzzi08}
\bibinfo{author}{S.~Zuzzi}, \bibinfo{author}{C.~Cametti},
  \bibinfo{author}{G.~Onori},
\newblock \bibinfo{title}{Polyion-induced aggregation of lipidic-coated solid
  polystyrene spheres: The many facets of complex formation in low-density
  colloidal suspensions},
\newblock \bibinfo{journal}{Langmuir} \bibinfo{volume}{24}
  (\bibinfo{year}{2008}) \bibinfo{pages}{6044--6049}.
\bibitem[{Milkova et~al.(2008)Milkova, Kamburova, Petkanchin, and
  Radeva}]{Radeva08}
\bibinfo{author}{V.~Milkova}, \bibinfo{author}{K.~Kamburova},
  \bibinfo{author}{I.~Petkanchin}, \bibinfo{author}{T.~Radeva},
\newblock \bibinfo{title}{Complexation of ferric oxide particles with pectins
  of different charge density},
\newblock \bibinfo{journal}{Langmuir} \bibinfo{volume}{24}
  (\bibinfo{year}{2008}) \bibinfo{pages}{9495--9499}.
\bibitem[{Kabanov et~al.(2000)Kabanov, Sergeyev, Pyshkina, Zinchenco, Zezin,
  Joosten, Brackman, and Yoshikawa}]{Kabanov00}
\bibinfo{author}{V.~Kabanov}, \bibinfo{author}{V.~Sergeyev},
  \bibinfo{author}{O.~Pyshkina}, \bibinfo{author}{A.~Zinchenco},
  \bibinfo{author}{A.~Zezin}, \bibinfo{author}{J.~Joosten},
  \bibinfo{author}{J.~Brackman}, \bibinfo{author}{K.~Yoshikawa},
\newblock \bibinfo{title}{Interpolyelectrolyte complexes formed by \text{DNA}
  and astramol poly(propylene imine) dendrimers},
\newblock \bibinfo{journal}{Macromolecules} \bibinfo{volume}{33}
  (\bibinfo{year}{2000}) \bibinfo{pages}{9587--9593}.
\bibitem[{Wang et~al.(2000)Wang, Kimura, Dubin, and Jaeger}]{Wang00}
\bibinfo{author}{Y.~Wang}, \bibinfo{author}{K.~Kimura}, \bibinfo{author}{P.~L.
  Dubin}, \bibinfo{author}{W.~Jaeger},
\newblock \bibinfo{title}{Polyelectrolyte-micelle coacervation: Effects of
  micelle surface charge density, polymer molecular weight, and
  polymer/surfactant ratio},
\newblock \bibinfo{journal}{Macromolecules} \bibinfo{volume}{33}
  (\bibinfo{year}{2000}) \bibinfo{pages}{3324--3331}.
\bibitem[{Sennato et~al.(2012)Sennato, Truzzolillo, and Bordi}]{SennatoSM2012}
\bibinfo{author}{S.~Sennato}, \bibinfo{author}{D.~Truzzolillo},
  \bibinfo{author}{F.~Bordi},
\newblock \bibinfo{title}{Aggregation and stability of
  polyelectrolyte-decorated liposome complexes in water--salt media},
\newblock \bibinfo{journal}{Soft Matter} \bibinfo{volume}{8}
  (\bibinfo{year}{2012}) \bibinfo{pages}{9384--9395}.
\bibitem[{Falk et~al.(2001)Falk, Lindman, Bengmark, Larsson, and
  Holmdahl}]{Falk01}
\bibinfo{author}{K.~Falk}, \bibinfo{author}{B.~Lindman},
  \bibinfo{author}{S.~Bengmark}, \bibinfo{author}{K.~Larsson},
  \bibinfo{author}{L.~Holmdahl},
\newblock \bibinfo{title}{Sodium polyacrylate potentiates the anti-adhesion
  effect of a cellulose-derived polymer},
\newblock \bibinfo{journal}{Biomaterials} \bibinfo{volume}{22}
  (\bibinfo{year}{2001}) \bibinfo{pages}{2185--90}.
\bibitem[{De~Giglio et~al.(2010)De~Giglio, Cafagna, Ricci, Sabbatini, Cometa,
  Ferretti, and Mattioli-Belmonte}]{DeGiglio10}
\bibinfo{author}{E.~De~Giglio}, \bibinfo{author}{D.~Cafagna},
  \bibinfo{author}{M.~A. Ricci}, \bibinfo{author}{L.~Sabbatini},
  \bibinfo{author}{S.~Cometa}, \bibinfo{author}{C.~Ferretti},
  \bibinfo{author}{M.~Mattioli-Belmonte},
\newblock \bibinfo{title}{Biocompatibility of poly(acrylic acid) thin coatings
  electro-synthesized onto \text{TiAlV}-based implants},
\newblock \bibinfo{journal}{J. Bioact. \& Compatible Polym.}
  \bibinfo{volume}{25} (\bibinfo{year}{2010}) \bibinfo{pages}{374--391}.
\bibitem[{Dhadwal et~al.(1991)Dhadwal, Ansari, and Meyer}]{Dhadwal91}
\bibinfo{author}{H.~S. Dhadwal}, \bibinfo{author}{R.~R. Ansari},
  \bibinfo{author}{W.~V. Meyer},
\newblock \bibinfo{title}{A fiber optic probe for particle sizing in
  concentrated suspensions},
\newblock \bibinfo{journal}{Rev. Sci. Instrum.} \bibinfo{volume}{62}
  (\bibinfo{year}{1991}) \bibinfo{pages}{2963--2968}.
\bibitem[{Provencher(1982)}]{Provencher82}
\bibinfo{author}{S.~Provencher},
\newblock \bibinfo{title}{A constrained regularization method for inverting
  data represented by linear algebraic or integral equations},
\newblock \bibinfo{journal}{Comput. Phys. Commun.} \bibinfo{volume}{27}
  (\bibinfo{year}{1982}) \bibinfo{pages}{213--227}.
\bibitem[{De~Vos et~al.(1996)De~Vos, Deriemaeker, and Finsy}]{Devos96}
\bibinfo{author}{C.~De~Vos}, \bibinfo{author}{L.~Deriemaeker},
  \bibinfo{author}{R.~Finsy},
\newblock \bibinfo{title}{Quantitative assessment of the conditioning of the
  inversion of quasi-elastic and static light scattering data for particle size
  distributions},
\newblock \bibinfo{journal}{Langmuir} \bibinfo{volume}{12}
  (\bibinfo{year}{1996}) \bibinfo{pages}{2630--2636}.
\bibitem[{Tscharnuter(2001)}]{Tscharnuter01}
\bibinfo{author}{W.~W. Tscharnuter},
\newblock \bibinfo{title}{Mobility measurements by phase analysis},
\newblock \bibinfo{journal}{Appl. Opt.} \bibinfo{volume}{40}
  (\bibinfo{year}{2001}) \bibinfo{pages}{3995--4003}.
\bibitem[{Hidalgo-Alvarez et~al.(1996)Hidalgo-Alvarez, Martin, Fernandez,
  Bastos, Martinez, and de~las Nieves}]{Hidalgo96}
\bibinfo{author}{R.~Hidalgo-Alvarez}, \bibinfo{author}{A.~Martin},
  \bibinfo{author}{A.~Fernandez}, \bibinfo{author}{D.~Bastos},
  \bibinfo{author}{F.~Martinez}, \bibinfo{author}{F.~de~las Nieves},
\newblock \bibinfo{title}{Electrokinetic properties, colloidal stability ad
  aggregation kinetics of polimer colloids},
\newblock \bibinfo{journal}{Adv. Colloid Interface Sci.} \bibinfo{volume}{67}
  (\bibinfo{year}{1996}) \bibinfo{pages}{1--118}.
\bibitem[{Truzzolillo et~al.(2009)Truzzolillo, Bordi, Sciortino, and
  Cametti}]{Truzzolillo09}
\bibinfo{author}{D.~Truzzolillo}, \bibinfo{author}{F.~Bordi},
  \bibinfo{author}{F.~Sciortino}, \bibinfo{author}{C.~Cametti},
\newblock \bibinfo{title}{Kinetic arrest in polyion-induced
  inhomogeneously-charged colloidal particle aggregation},
\newblock \bibinfo{journal}{Eur. Phys. J. E} \bibinfo{volume}{29}
  (\bibinfo{year}{2009}) \bibinfo{pages}{229--237}.
\bibitem[{Tadmor(2001)}]{Tadmor01}
\bibinfo{author}{R.~Tadmor},
\newblock \bibinfo{title}{The \text{L}ondon-van der \text{W}aals interaction
  energy between objects of various geometries},
\newblock \bibinfo{journal}{J. Phys.: Cond. Matt.} \bibinfo{volume}{13}
  (\bibinfo{year}{2001}) \bibinfo{pages}{L195}.
\bibitem[{Hogg et~al.(1966)Hogg, Healy, and Fuerstenau}]{Hogg66}
\bibinfo{author}{R.~Hogg}, \bibinfo{author}{T.~W. Healy},
  \bibinfo{author}{D.~W. Fuerstenau},
\newblock \bibinfo{title}{Mutual coagulation of colloidal dispersions},
\newblock \bibinfo{journal}{Trans. Faraday Soc.} \bibinfo{volume}{62}
  (\bibinfo{year}{1966}) \bibinfo{pages}{1638--1651}.
\bibitem[{Derjaguin(1934)}]{Derjaguin34}
\bibinfo{author}{B.~V. Derjaguin},
\newblock \bibinfo{title}{Analysis of friction and adhesion. \text{IV. T}he
  theory of the adhesion of small particles},
\newblock \bibinfo{journal}{Kolloid Z.} \bibinfo{volume}{69}
  (\bibinfo{year}{1934}) \bibinfo{pages}{155--164}.
\bibitem[{Israelachvili(1985)}]{Israelachvili85}
\bibinfo{author}{J.~N. Israelachvili}, \bibinfo{title}{Intermolecular and
  surface forces}, \bibinfo{publisher}{Academic Press},
  \bibinfo{address}{London}, \bibinfo{year}{1985}.
\bibitem[{Rentsch et~al.(2006)Rentsch, Pericet-Camara, Papastavrou, and
  Borkovec}]{Borkovec06}
\bibinfo{author}{S.~Rentsch}, \bibinfo{author}{R.~Pericet-Camara},
  \bibinfo{author}{G.~Papastavrou}, \bibinfo{author}{M.~Borkovec},
\newblock \bibinfo{title}{Probing the validity of the derjaguin approximation
  for heterogeneous colloidal particles},
\newblock \bibinfo{journal}{Phys. Chem. Chem. Phys.} \bibinfo{volume}{8}
  (\bibinfo{year}{2006}) \bibinfo{pages}{2531--2538}.
\bibitem[{Dzubiella et~al.(2003)Dzubiella, Moreira, and Pincus}]{Dzubiella03}
\bibinfo{author}{J.~Dzubiella}, \bibinfo{author}{A.~G. Moreira},
  \bibinfo{author}{P.~A. Pincus},
\newblock \bibinfo{title}{Polyelectrolyte-colloid complexes: Polarizability and
  effective interaction},
\newblock \bibinfo{journal}{Macromolecules} \bibinfo{volume}{36}
  (\bibinfo{year}{2003}) \bibinfo{pages}{1741--1752}.
\bibitem[{Granfeldt et~al.(1991)Granfeldt, Joansson, and
  Woodward}]{Granfeldt91}
\bibinfo{author}{M.~K. Granfeldt}, \bibinfo{author}{B.~Joansson},
  \bibinfo{author}{C.~E. Woodward},
\newblock \bibinfo{title}{A \text{M}onte \text{C}arlo simulation study of the
  interaction between charged colloids carrying adsorbed polyelectrolytes},
\newblock \bibinfo{journal}{J. Phys. Chem.} \bibinfo{volume}{95}
  (\bibinfo{year}{1991}) \bibinfo{pages}{4819--4826}.
\bibitem[{Sennato et~al.(2009)Sennato, Truzzolillo, Bordi, Sciortino, and
  Cametti}]{Sennato09}
\bibinfo{author}{S.~Sennato}, \bibinfo{author}{D.~Truzzolillo},
  \bibinfo{author}{F.~Bordi}, \bibinfo{author}{F.~Sciortino},
  \bibinfo{author}{C.~Cametti},
\newblock \bibinfo{title}{Colloidal particle aggregates induced by particle
  surface charge heterogeneity},
\newblock \bibinfo{journal}{Coll. Surf. A} \bibinfo{volume}{343}
  (\bibinfo{year}{2009}) \bibinfo{pages}{34--42}.
\bibitem[{Wang et~al.(1999)Wang, Kimura, Huang, and Dubin}]{Wang99}
\bibinfo{author}{Y.~Wang}, \bibinfo{author}{K.~Kimura},
  \bibinfo{author}{Q.~Huang}, \bibinfo{author}{P.~L. Dubin},
\newblock \bibinfo{title}{Effect of salt on polyelectrolyte-micelle
  coacervation},
\newblock \bibinfo{journal}{Macromolecules} \bibinfo{volume}{32}
  (\bibinfo{year}{1999}) \bibinfo{pages}{7128--7134}.
\bibitem[{Cevc(1993)}]{Cevc93}
\bibinfo{author}{G.~Cevc},
\newblock \bibinfo{title}{Electrostatic characterization of liposomes},
\newblock \bibinfo{journal}{Chem. Phys. Lipids} \bibinfo{volume}{64}
  (\bibinfo{year}{1993}) \bibinfo{pages}{163--186}.
\bibitem[{Muthukumar(1987)}]{Muthukumar87}
\bibinfo{author}{M.~Muthukumar},
\newblock \bibinfo{title}{Adsorption of a polyelectrolyte chain to a charged
  surface},
\newblock \bibinfo{journal}{J. Chem. Phys.} \bibinfo{volume}{86}
  (\bibinfo{year}{1987}) \bibinfo{pages}{7230--7235}.
\bibitem[{Sennato et~al.(2001)Sennato, Bordi, and Cametti}]{SennatoEPL04}
\bibinfo{author}{S.~Sennato}, \bibinfo{author}{F.~Bordi},
  \bibinfo{author}{C.~Cametti},
\newblock \bibinfo{title}{Correlated adsorption of polyelectrolytes in the
  "charge inversion" of colloidal particles},
\newblock \bibinfo{journal}{Europhys. Lett.} \bibinfo{volume}{68}
  (\bibinfo{year}{2001}) \bibinfo{pages}{296--302}.
\bibitem[{Sabin et~al.(2006)Sabin, Prieto, Ruso, Hidalgo-Alvarez, and
  Sarmiento}]{SabinEPJE}
\bibinfo{author}{J.~Sabin}, \bibinfo{author}{G.~Prieto},
  \bibinfo{author}{J.~Ruso}, \bibinfo{author}{R.~Hidalgo-Alvarez},
  \bibinfo{author}{F.~Sarmiento},
\newblock \bibinfo{title}{Size and stability of liposomes: A possible role of
  hydration and osmotic forces},
\newblock \bibinfo{journal}{Eur. Phys. J. E} \bibinfo{volume}{20}
  (\bibinfo{year}{2006}) \bibinfo{pages}{401--8}.
\bibitem[{Hirsch-Lerner et~al.(2005)Hirsch-Lerner, Zhang, Eliyahu, Ferrari,
  Wheeler, and Barenholz}]{Hirsch-Lerner05}
\bibinfo{author}{D.~Hirsch-Lerner}, \bibinfo{author}{M.~Zhang},
  \bibinfo{author}{H.~Eliyahu}, \bibinfo{author}{M.~E. Ferrari},
  \bibinfo{author}{C.~J. Wheeler}, \bibinfo{author}{Y.~Barenholz},
\newblock \bibinfo{title}{Effect of 'helper lipid' on lipoplex electrostatics},
\newblock \bibinfo{journal}{Biochim. Biophys. Acta} \bibinfo{volume}{1714}
  (\bibinfo{year}{2005}) \bibinfo{pages}{71--84}.
\bibitem[{Zuidam and Barenholz(1999)}]{zuidam1999characterization}
\bibinfo{author}{N.~Zuidam}, \bibinfo{author}{Y.~Barenholz},
\newblock \bibinfo{title}{Characterization of \text{DNA}-lipid complexes
  commonly used for gene delivery},
\newblock \bibinfo{journal}{Int. J. Pharm.} \bibinfo{volume}{183}
  (\bibinfo{year}{1999}) \bibinfo{pages}{43--46}.
\bibitem[{Koltover et~al.(1999)Koltover, Salditt, and Safinya}]{Safinya99}
\bibinfo{author}{I.~Koltover}, \bibinfo{author}{T.~Salditt},
  \bibinfo{author}{C.~R. Safinya},
\newblock \bibinfo{title}{Phase diagram, stability, and overcharging of
  lamellar cationic lipid-\text{DNA} self-assembled complexes},
\newblock \bibinfo{journal}{Biophys. J.} \bibinfo{volume}{77}
  (\bibinfo{year}{1999}) \bibinfo{pages}{915--924}.
\bibitem[{Bordi et~al.(2008)Bordi, Cametti, Di~Venanzio, Sennato, and
  Zuzzi}]{Bordi08}
\bibinfo{author}{F.~Bordi}, \bibinfo{author}{C.~Cametti},
  \bibinfo{author}{C.~Di~Venanzio}, \bibinfo{author}{S.~Sennato},
  \bibinfo{author}{S.~Zuzzi},
\newblock \bibinfo{title}{Influence of temperature on microdomain organization
  of mixed cationic-zwitterionic lipidic monolayers at the air-water
  interface},
\newblock \bibinfo{journal}{Coll. Surf. B} \bibinfo{volume}{61}
  (\bibinfo{year}{2008}) \bibinfo{pages}{304--310}.
\bibitem[{Cinelli et~al.(2007)Cinelli, Onori, Zuzzi, Bordi, Cametti, Sennato,
  and Diociaiuti}]{Cinelli07}
\bibinfo{author}{S.~Cinelli}, \bibinfo{author}{G.~Onori},
  \bibinfo{author}{S.~Zuzzi}, \bibinfo{author}{F.~Bordi},
  \bibinfo{author}{C.~Cametti}, \bibinfo{author}{S.~Sennato},
  \bibinfo{author}{M.~Diociaiuti},
\newblock \bibinfo{title}{Properties of mixed \text{DOTAP-DPPC} bilayer
  membranes as reported by differential scanning calorimetry and dynamic light
  scattering measurements},
\newblock \bibinfo{journal}{J. Phys. Chem. B} \bibinfo{volume}{111}
  (\bibinfo{year}{2007}) \bibinfo{pages}{10032--10039}.
\bibitem[{Yaroslavov et~al.(2011)Yaroslavov, Efimova, Sybachin, Izumrudov, and
  Potemkin}]{Yaroslavov11}
\bibinfo{author}{A.~A. Yaroslavov}, \bibinfo{author}{A.~A. Efimova},
  \bibinfo{author}{A.~A. Sybachin}, \bibinfo{author}{V.~A. Izumrudov, V.
  A.~Samoshin}, \bibinfo{author}{V.~A. Potemkin},
\newblock \bibinfo{title}{Stability of anionic liposome-cationic polymer
  complexes in water-salt media},
\newblock \bibinfo{journal}{Colloid J.} \bibinfo{volume}{73}
  (\bibinfo{year}{2011}) \bibinfo{pages}{430--435}.
\bibitem[{Wiegel(1977)}]{Wiegel77}
\bibinfo{author}{F.~W. Wiegel},
\newblock \bibinfo{title}{Adsorption of a macromolecule to a charged surface},
\newblock \bibinfo{journal}{J. Phys. A: Math. Gen.} \bibinfo{volume}{10}
  (\bibinfo{year}{1977}) \bibinfo{pages}{299--303}.
\bibitem[{von Goeler and Muthukumar(1994)}]{vonGoeler94}
\bibinfo{author}{F.~von Goeler}, \bibinfo{author}{M.~Muthukumar},
\newblock \bibinfo{title}{Adsorption of polyelectrolytes onto curved
  surfaces.},
\newblock \bibinfo{journal}{J. Chem. Phys.} \bibinfo{volume}{100}
  (\bibinfo{year}{1994}) \bibinfo{pages}{7796--7802}.
\bibitem[{Haronska et~al.(1998)Haronska, Vilgis, Grottenm\"{u}ller, and
  Schmidt}]{Haronska98}
\bibinfo{author}{P.~Haronska}, \bibinfo{author}{T.~A. Vilgis},
  \bibinfo{author}{R.~Grottenm\"{u}ller}, \bibinfo{author}{M.~Schmidt},
\newblock \bibinfo{title}{Adsorption of polymer chains onto charged spheres:
  Experiment and theory},
\newblock \bibinfo{journal}{Macromol. Theory Simul.} \bibinfo{volume}{7}
  (\bibinfo{year}{1998}) \bibinfo{pages}{241}.
\bibitem[{Hansupalak and Santore(2003)}]{Hansupalak03}
\bibinfo{author}{N.~Hansupalak}, \bibinfo{author}{M.~M. Santore},
\newblock \bibinfo{title}{Sharp polyelectrolyte adsorption cutoff induced by
  monovalent salt},
\newblock \bibinfo{journal}{Langmuir} \bibinfo{volume}{19}
  (\bibinfo{year}{2003}) \bibinfo{pages}{7423--7426}.
\bibitem[{Zhang et~al.(2000)Zhang, Ohbu, and Dubin}]{Zhang00}
\bibinfo{author}{H.~Zhang}, \bibinfo{author}{K.~Ohbu}, \bibinfo{author}{P.~L.
  Dubin},
\newblock \bibinfo{title}{Binding of carboxy-terminated anionic/nonionic mixed
  micelles to a strong polycation: Critical conditions for complex formation},
\newblock \bibinfo{journal}{Langmuir} \bibinfo{volume}{16}
  (\bibinfo{year}{2000}) \bibinfo{pages}{9082--9086}.
\bibitem[{Mylonas et~al.(1999)Mylonas, Staikos, and Ullner}]{Mylonas99}
\bibinfo{author}{Y.~Mylonas}, \bibinfo{author}{G.~Staikos},
  \bibinfo{author}{M.~Ullner},
\newblock \bibinfo{title}{Chain conformation and intermolecular interaction of
  partially neutralized poly(acrylic acid) in dilute aqueous solutions},
\newblock \bibinfo{journal}{Polymer} \bibinfo{volume}{40}
  (\bibinfo{year}{1999}) \bibinfo{pages}{6841--6847}.
\bibitem[{Petrache et~al.(2004)Petrache, Tristram-Nagle, Gawrisch, Harries,
  Parsegian, and Nagle}]{Petrache04}
\bibinfo{author}{H.~L. Petrache}, \bibinfo{author}{S.~Tristram-Nagle},
  \bibinfo{author}{K.~Gawrisch}, \bibinfo{author}{D.~Harries},
  \bibinfo{author}{V.~A. Parsegian}, \bibinfo{author}{J.~Nagle},
\newblock \bibinfo{title}{Structure and fluctuations of charged
  phosphatidylserine bilayers in the absence of salt},
\newblock \bibinfo{journal}{Biophys. J.} \bibinfo{volume}{86}
  (\bibinfo{year}{2004}) \bibinfo{pages}{1574--1586}.
\bibitem[{Quesada-Perez et~al.(2002)Quesada-Perez, Callejas-Fernandez, and
  Hidalgo-Alvarez}]{Quesada-Perez02}
\bibinfo{author}{M.~Quesada-Perez}, \bibinfo{author}{J.~Callejas-Fernandez},
  \bibinfo{author}{A.~R. Hidalgo-Alvarez},
\newblock \bibinfo{title}{Interaction potentials, structural ordering and
  effective charges in dispersions of charged colloidal particles},
\newblock \bibinfo{journal}{Adv. Colloid Interf. Sci.} \bibinfo{volume}{95}
  (\bibinfo{year}{2002}) \bibinfo{pages}{295--315}.
\bibitem[{Haro-P{\'e}rez et~al.(2003)Haro-P{\'e}rez, Quesada-P{\'e}rez,
  Callejas-Fern{\'a}ndez, Casals, Estelrich, and
  Hidalgo-{\'A}lvarez}]{haro2003}
\bibinfo{author}{C.~Haro-P{\'e}rez}, \bibinfo{author}{M.~Quesada-P{\'e}rez},
  \bibinfo{author}{J.~Callejas-Fern{\'a}ndez}, \bibinfo{author}{E.~Casals},
  \bibinfo{author}{J.~Estelrich}, \bibinfo{author}{R.~Hidalgo-{\'A}lvarez},
\newblock \bibinfo{title}{Liquidlike structures in dilute suspensions of
  charged liposomes},
\newblock \bibinfo{journal}{J. Chem. Phys.} \bibinfo{volume}{118}
  (\bibinfo{year}{2003}) \bibinfo{pages}{5167--5173}.
\bibitem[{Bordi et~al.(2006)Bordi, Cametti, Sennato, Paoli, and
  Marianecci}]{Bordi06e}
\bibinfo{author}{F.~Bordi}, \bibinfo{author}{C.~Cametti},
  \bibinfo{author}{S.~Sennato}, \bibinfo{author}{B.~Paoli},
  \bibinfo{author}{C.~Marianecci},
\newblock \bibinfo{title}{Charge renormalization in planar and spherical
  charged lipidic aqueous interfaces.},
\newblock \bibinfo{journal}{J. Phys. Chem. B} \bibinfo{volume}{110}
  (\bibinfo{year}{2006}) \bibinfo{pages}{4808--4814}.
\bibitem[{Aubouy et~al.(2003)Aubouy, Trizac, and Bocquet}]{Auboy03}
\bibinfo{author}{M.~Aubouy}, \bibinfo{author}{E.~Trizac},
  \bibinfo{author}{L.~Bocquet},
\newblock \bibinfo{title}{Effective charge versus bare charge: an analytical
  estimate for colloids in the infinite dilution limit},
\newblock \bibinfo{journal}{J. Phys. A: Math. Gen.} \bibinfo{volume}{36}
  (\bibinfo{year}{2003}) \bibinfo{pages}{5835--5840}.
\bibitem[{Bordi et~al.(2004)Bordi, Cametti, and Colby}]{Bordi04b}
\bibinfo{author}{F.~Bordi}, \bibinfo{author}{C.~Cametti},
  \bibinfo{author}{R.~H. Colby},
\newblock \bibinfo{title}{Dielectric spectroscopy and conductivity of
  polyelectrolyte solutions},
\newblock \bibinfo{journal}{J. Phys.: Cond. Matt.,} \bibinfo{volume}{16}
  (\bibinfo{year}{2004}) \bibinfo{pages}{R1423--R1463}.
\bibitem[{Bordi et~al.(2002)Bordi, Colby, Cametti, De~Lorenzo, and
  Gili}]{Bordi02}
\bibinfo{author}{F.~Bordi}, \bibinfo{author}{R.~H. Colby},
  \bibinfo{author}{C.~Cametti}, \bibinfo{author}{L.~De~Lorenzo},
  \bibinfo{author}{T.~Gili},
\newblock \bibinfo{title}{Electrical conductivity of polyelectrolyte solutions
  in the semidilute and concentrated regime: The role of counterion
  condensation},
\newblock \bibinfo{journal}{J. Phys. Chem. B} \bibinfo{volume}{106}
  (\bibinfo{year}{2002}) \bibinfo{pages}{6887--6893}.
\bibitem[{Fleck and von Gr{\"u}nberg(2001)}]{fleck2001counterion}
\bibinfo{author}{C.~Fleck}, \bibinfo{author}{H.~von Gr{\"u}nberg},
\newblock \bibinfo{title}{Counterion evaporation},
\newblock \bibinfo{journal}{Phys. Rev. E} \bibinfo{volume}{63}
  (\bibinfo{year}{2001}) \bibinfo{pages}{061804}.
\bibitem[{Park et~al.(1999)Park, Bruinsma, and Gelbart}]{park1999spontaneous}
\bibinfo{author}{S.~Park}, \bibinfo{author}{R.~Bruinsma},
  \bibinfo{author}{W.~Gelbart},
\newblock \bibinfo{title}{Spontaneous overcharging of macro-ion complexes},
\newblock \bibinfo{journal}{EPL (Europhysics Letters)} \bibinfo{volume}{46}
  (\bibinfo{year}{1999}) \bibinfo{pages}{454}.
\bibitem[{Sens and Joanny(2000)}]{sens2000counterion}
\bibinfo{author}{P.~Sens}, \bibinfo{author}{J.-F. Joanny},
\newblock \bibinfo{title}{Counterion release and electrostatic adsorption},
\newblock \bibinfo{journal}{Phys. Rev. Lett.} \bibinfo{volume}{84}
  (\bibinfo{year}{2000}) \bibinfo{pages}{4862}.
\bibitem[{Dobrynin et~al.(2000)Dobrynin, Deshkovski, and
  Rubinstein}]{dobrynin2000adsorption}
\bibinfo{author}{A.~V. Dobrynin}, \bibinfo{author}{A.~Deshkovski},
  \bibinfo{author}{M.~Rubinstein},
\newblock \bibinfo{title}{Adsorption of polyelectrolytes at an oppositely
  charged surface},
\newblock \bibinfo{journal}{Phys. Rev. Lett.} \bibinfo{volume}{84}
  (\bibinfo{year}{2000}) \bibinfo{pages}{3101}.
\bibitem[{Bordi et~al.(2003)Bordi, Cametti, and Gili}]{Bordi03pre}
\bibinfo{author}{F.~Bordi}, \bibinfo{author}{C.~Cametti},
  \bibinfo{author}{T.~Gili},
\newblock \bibinfo{title}{Electrical conductivity of polyelectrolyte solutions
  in the presence of added salt: The role of the solvent quality factor in
  light of a scaling approach},
\newblock \bibinfo{journal}{Phys. Rev. E} \bibinfo{volume}{68}
  (\bibinfo{year}{2003}) \bibinfo{pages}{011805}.
\bibitem[{Bordi et~al.(2006)Bordi, Cametti, Sennato, and Viscomi}]{Bordi06}
\bibinfo{author}{F.~Bordi}, \bibinfo{author}{C.~Cametti},
  \bibinfo{author}{S.~Sennato}, \bibinfo{author}{D.~Viscomi},
\newblock \bibinfo{title}{Counterion release in overcharging of
  polyion-liposome complexes},
\newblock \bibinfo{journal}{Phys. Rev. E.} \bibinfo{volume}{74}
  (\bibinfo{year}{2006}) \bibinfo{pages}{030402R/1--4}.
\bibitem[{Monopoli et~al.(2011)Monopoli, Walczyk, Campbell, Elia, Lynch,
  Baldelli~Bombelli, and Dawson}]{monopoli2011physical}
\bibinfo{author}{M.~P. Monopoli}, \bibinfo{author}{D.~Walczyk},
  \bibinfo{author}{A.~Campbell}, \bibinfo{author}{G.~Elia},
  \bibinfo{author}{I.~Lynch}, \bibinfo{author}{F.~Baldelli~Bombelli},
  \bibinfo{author}{K.~A. Dawson},
\newblock \bibinfo{title}{Physical- chemical aspects of protein corona:
  relevance to in vitro and in vivo biological impacts of nanoparticles},
\newblock \bibinfo{journal}{JACS} \bibinfo{volume}{133} (\bibinfo{year}{2011})
  \bibinfo{pages}{2525--2534}.
\bibitem[{Wang et~al.(2011)Wang, Jensen, Jensen, Shipovskov, Balakrishnan,
  Otzen, Pedersen, Besenbacher, and Sutherland}]{wang2011soft}
\bibinfo{author}{J.~Wang}, \bibinfo{author}{U.~B. Jensen},
  \bibinfo{author}{G.~V. Jensen}, \bibinfo{author}{S.~Shipovskov},
  \bibinfo{author}{V.~S. Balakrishnan}, \bibinfo{author}{D.~Otzen},
  \bibinfo{author}{J.~S. Pedersen}, \bibinfo{author}{F.~Besenbacher},
  \bibinfo{author}{D.~S. Sutherland},
\newblock \bibinfo{title}{Soft interactions at nanoparticles alter protein
  function and conformation in a size dependent manner},
\newblock \bibinfo{journal}{Nano Letters} \bibinfo{volume}{11}
  (\bibinfo{year}{2011}) \bibinfo{pages}{4985--4991}.
\bibitem[{Kah et~al.(2012)Kah, Chen, Zubieta, and
  Hamad-Schifferli}]{kah2012exploiting}
\bibinfo{author}{J.~C.~Y. Kah}, \bibinfo{author}{J.~Chen},
  \bibinfo{author}{A.~Zubieta}, \bibinfo{author}{K.~Hamad-Schifferli},
\newblock \bibinfo{title}{Exploiting the protein corona around gold nanorods
  for loading and triggered release},
\newblock \bibinfo{journal}{ACS Nano} \bibinfo{volume}{6}
  (\bibinfo{year}{2012}) \bibinfo{pages}{6730--6740}.

\end{thebibliography}
\bibliographystyle{model1-num-names}

\end{document}